\documentclass[fleqn,usenatbib]{mnras}

\usepackage[T1]{fontenc}
\usepackage{ae,aecompl}

\usepackage{bm}
\usepackage{fix-cm}     
\usepackage{graphicx}	
\usepackage[export]{adjustbox} 
\usepackage{amsmath}	
\usepackage{amssymb}	
\usepackage{adjustbox}
\usepackage{comment}
\usepackage{siunitx}
\usepackage{float}
\usepackage[table,xcdraw]{xcolor}
\usepackage{xspace}
\usepackage{etoolbox}
\usepackage{multirow}
\usepackage{tcolorbox}

\usepackage{newtxtext,newtxmath}

\title{Gas in Globular Clusters I: Gas Retention and Its Possible Consequences}

\author[A.~Bobrick et al.]{Alexey~Bobrick$^{1,2,3}$\thanks{alexey.bobrick@monash.edu}, Melvyn~B.~Davies$^4$, Hagai~B.~Perets$^3$\\ 
$^1$ School of Physics and Astronomy, Monash University, Clayton, VIC 3800, Australia\\
$^2$ OzGrav: Australian Research Council Centre of Excellence for Gravitational Wave Discovery, Clayton, VIC 3800, Australia\\
$^3$ Technion --- Israel Institute of Technology, Physics Department, Haifa, Israel 32000\\
$^4$ Centre for Mathematical Sciences, Lund University, Box 118, SE 221-00 Lund, Sweden\\
}
\date{Accepted XXX. Received YYY; in original form ZZZ}

\pubyear{2025}

\begin{document}
\label{firstpage}
\pagerange{\pageref{firstpage}--\pageref{lastpage}}
\maketitle

\begin{abstract}
Globular clusters host complex stellar populations whose chemical signatures suggest early ($3\times10^6$~--~$10^9\,\mathrm{yr}$) retention and reprocessing of stellar ejecta, yet direct evidence for intracluster gas is lacking. Here we present a unified theoretical framework for the evolution of gas in young globular clusters, and its implications for the production of multiple stellar populations. We show that low-velocity AGB winds are gravitationally retained in clusters more massive than a few\,$\times10^5\,\text{M}_\odot$. In addition, AGB winds in such clusters collide with each other and the previously retained winds, triggering a rapid `switch' to efficient gas retention. Expected gas retention fractions agree well with the observed second population fractions in Galactic globular clusters. Furthermore, the accumulated gas cannot form new stars because protostellar cores are disrupted by encounters with pre-existing stars. Instead, the gas is accreted onto pre-existing main-sequence stars and compact objects. Time-dependent core-halo models indicate that compact objects can grow and collapse within $\sim10^8$~--~$10^9\,\mathrm{yr}$, while lower-mass main-sequence stars can be `rejuvenated' into the $4$~--~$6\,\text{M}_\odot$ range required to reproduce key abundance patterns. Therefore, in our model, the multiple populations will be found in sufficiently massive clusters, with the second-population stars being formed from the inner subset of first-population stars that accreted large fractions of their mass from the AGB-processed retained gas. Finally, we argue that a combination of feedback processes, including accretion luminosity onto compact objects, novae, pulsar winds, and binary supernovae, will clear the gas by $\lesssim10^9\,\mathrm{yr}$, thus reproducing the gas-poor conditions observed for present-day clusters.
\end{abstract}





\section{Introduction}

\subsection{Context and motivation}

Globular clusters (GCs) are among the oldest bound stellar systems in the Galaxy, containing $10^4$~--~$10^7$ stars within a few parsec radii, giving insights into early star formation in the Universe \citep{Harris96}.  Despite their age and apparent simplicity, GCs are factories of stellar exotica. They host surprisingly enhanced and diverse stellar populations, including low-mass X-ray binaries (LMXBs), millisecond pulsars (MSPs), gravitational-wave (GW) progenitors, and (possibly) intermediate-mass black holes (IMBHs) \citep{Verbunt1987, Pooley03, Kremer18}.  Moreover, high-precision photometry and spectroscopy have revealed that most GCs harbour multiple chemically distinct stellar populations (MPs), challenging the long‐standing paradigm of GCs as simple, single-population systems \citep{Milone22}.  

The existence of these multiple populations, characterised by light element abundance variations, such as the Na-O anticorrelation, and discrete main-sequence splits, implies that some fraction of processed stellar ejecta must have been retained, mixed, and incorporated into later-forming stars within the first few $10^8$~--~$10^9\,\text{yr}$ of a cluster's life \citep{Conroy2011,Bastian13}.  Yet, direct searches for intracluster gas in present-day GCs yield stringent upper limits of $M_{\text{gas,today}}\ll1\,{\text{M}}_\odot$ per cluster, based on H\,I and OH maser surveys, dust emission, and pulsar dispersion measures \citep{Cohen1979, Lynch1989, vanLoon2009, Abbate2018}.  
In this work (Paper~I of a series), we develop a self‐consistent framework for how low-velocity stellar winds can be retained in sufficiently massive and compact GCs, how the retained gas evolves and interacts with the cluster stellar population and accretes onto it, and how feedback processes eventually clear the gas.

\subsection{Chemical signatures of multiple populations}

High‐precision photometry and spectroscopy have shown that the majority of Milky Way GCs host two or more chemically distinct stellar subpopulations \citep{Milone22}. One of the populations, called the first population, is chemically consistent with field stars. In contrast, the other one, called the second population, shows light element abundance variations, most notably Na-O and Mg-Al anticorrelations, among red giants, often showing Na enhancements of up to $\sim0.8\,$dex and corresponding O depletions \citep{Carretta09, Carretta12}.  These patterns extend down to the main sequence and subgiant branch, implying that the anomalies were inherited before or during star formation rather than by later stellar evolutionary mixing \citep{Gratton01, Gratton12}.  

Photometrically, narrow‐band blue and ultraviolet filters have been used to separate the first‐ and second‐population stars cleanly using so‐called `chromosome maps', showing discrete splits in the main sequence (e.g., in $\omega$ Cen and NGC 2808) and subgiant and red giant branches \citep{Piotto07, Milone08, Latour18}. Thanks to the large photometric samples, it was established that second‐population stars tend to be more centrally concentrated and exhibit lower binary fractions, consistent with formation from centrally retained gas \citep{Lardo11, Milone12, Lucatello15}.  

Quantitatively, the fraction of second‐population stars ($f_{\text{2P}}$) varies from $\sim20\%$ in low-mass clusters to $\gtrsim80\%$ in the most massive systems, correlating most strongly with the cluster’s inferred initial mass rather than its present-day mass or metallicity \citep{Milone2017, Baumgardt2018,Mas+21}. Reproducing both the high $f_{\text{2P}}$ and the discrete nature of the subpopulations requires that a reservoir of enriched gas amounting to $\gtrsim0.1$~--~$0.3\,M_{\text{init}}$ of the cluster mass remained bound after the initial phase of star formation, mixed with processed ejecta from intermediate‐mass stars ($M_\star\gtrsim 4\,\text{M}_\odot$) or possibly young very massive stars \cite{Decressin07} or binaries \citep{DeMink09}, and possibly formed new stars within $\sim10^8$~--~$10^9\,\text{yr}$ \citep{DErcole2008, Conroy2011, Bastian13}. Or, as suggested previously \citep{Perets2022}, and as we suggest here, the reservoir of gas got accreted on existing stars, producing a new population of stars composed of a mix between the original stellar material and newly accreted material from the retained AGB-processed gas in the cluster. A key aim of this work is to show how such a reservoir can naturally arise from gravitational retention of stellar winds, as well as wind collisions, and how its time-dependent evolution determines the observed first and second population fractions and spatial distributions.

\subsection{Observational limits on gas in globular clusters}

Direct searches for neutral and ionised gas in Galactic GCs have consistently returned non-detections or extremely low upper limits. Early H\,I and OH maser surveys constrained cold gas masses to $\lesssim10$~--~$100\,\text{M}_\odot$ per cluster \citep{Kerr1972, Knapp1973b, Cohen1979}, while more sensitive pointed observations of several nearby clusters pushed these limits down to a few $\text{M}_\odot$ or less \citep{Lynch1989, vanLoon2006}. Infrared measurements of dust emission in M15 revealed only $9\times10^{-4}\,{\text{M}}_\odot$ of dust, implying $\ll1\,\text{M}_\odot$ of associated gas \citep{Boyer2006}. Most recently, pulsar dispersion-measure studies in 47 Tuc placed a $3\sigma$ upper limit of $0.023\pm0.005\,\text{M}_\odot$ of ionised gas within the central parsec \citep{Abbate2018}.

Comparable non‐detections have been found in young massive clusters (ages $\lesssim1$~--~$2\,\text{Gyr}$) in the Magellanic Clouds, where both maser and dust searches limit gas masses to $\ll1\%$ of the cluster mass \citep{Bastian2014}. Moreover, photometric and spectroscopic surveys of these younger clusters reveal no evidence for discrete light-element abundance spreads or split photometric sequences, possibly indicating that multiple populations do not arise on timescales shorter than $\sim2\,\text{Gyr}$ \citep{Bastian2016, CabreraZiri2016}.

These stringent constraints contrast sharply with the much larger amount of gas that must have been retained early on to account for the observed second-population fractions in old GCs. Moreover, the present lack of gas cannot be ascribed solely to external stripping: while ram-pressure stripping during Galactic plane crossings can remove $10^2$~--~$10^3\,\text{M}_\odot$ every $\sim10^8\,\text{yr}$ \citep{Roberts1988}, the recurrence timescale is longer than the rapid stellar wind injection and recycling timescales, thus requiring an efficient internal removal mechanism as well.

\subsection{Theoretical challenges}

Any successful explanation for early gas retention and multiple populations in globular clusters must satisfy several stringent requirements:

\begin{enumerate}
  \item \textbf{Retention versus escape.}  
    Stellar winds from AGB stars have typical speeds $v_{\rm wind}\sim5$~--~$20\,\text{km}\,\text{s}^{-1}$ \citep{Ramstedt2008,Hoefner2018}, while cluster escape speeds scale as  
    \begin{equation}
        v_{\text{esc}}\simeq\sqrt{\frac{2GM_{\text{GC}}}{R_{\text{ hm}}}}\,.  
    \end{equation}
    Clusters with $v_{\text{esc}}\ll v_{\text{wind}}$ cannot gravitationally trap ejecta, whereas those with $v_{\text{esc}}\gtrsim v_{\text{wind}}$ may gravitationally retain a large fraction of the mass lost (Section~\ref{sec:GasRet}).

  \item \textbf{Collisionless versus collisional stellar winds.}
    In low-density environments, gravitationally unbound wind ejecta behave effectively as if they are collisionless.  However, in dense cluster cores, frequent wind-wind collisions may occur, thermally releasing their kinetic energy and slowing down the outflow, greatly enhancing retention (Section~\ref{sec:Winds}), which has been neglected in many previous analytic treatments \citep[e.g.][]{Krause2016}.

  \item \textbf{Fate of retained gas.}  
    Dense intracluster medium may cool and fragment, potentially forming new stars \citep{DErcole2008}. However, contracting protostellar cores even near the half-mass radius in globular clusters are subject to disruptive encounters on timescales $t_{\text{enc}}\sim 10^3-10^4\, \text{yr}$,
    preventing collapse and instead directing the gas into Bondi-Hoyle-type accretion flows (Section~\ref{sec:GasFate}).

  \item \textbf{Accretion and feedback.}  
    Retained gas with density $\rho$ and sound speed $c_s$ can be accreted onto main-sequence stars and compact objects with masses $M_\star$ and velocities $v$ at the Bondi-Hoyle rate $\dot M_{\text{B-H}}\propto M_\star^2/(c^2_s+v^2)^{3/2}$ \citep{Edgar2004}, thus increasing the stellar mass and releasing energy via accretion luminosity, novae, and, possibly, supernovae that may ultimately expel the gas \citep{Moore2011,Winter2023}.

  \item \textbf{Timescale consistency.}  
    All relevant processes, including stellar wind production, wind-wind collisions, protostellar core disruption, stellar accretion growth, and feedback gas clearing, must operate within $\sim10^8$~--~$10^9\,\text{yr}$ to match the epoch of second-population star formation and the present-day absence of gas \citep{Conroy2011,Naiman2018}.
\end{enumerate}

Although previous models have addressed individual aspects, such as gravitational potential retention of stars \citep{Krause2016}, stellar wind retention \cite{Mas+21}, ram-pressure effects \citep{Roberts1988}, or accretion feedback \citep{Moore2011}, a coherent, time-dependent theory that integrates wind physics, cluster dynamics, and stellar encounters has been lacking. The goal of this study is to fill this gap by developing a unified, self-consistent framework, which is presented in the following sections.

\subsection{Scope and outline of this work}

In this paper (Paper~I), we present a unified, time-dependent framework that connects stellar wind physics, cluster dynamics, and the origin of multiple stellar populations:

\begin{itemize}
  \item In Section~\ref{sec:Winds}, we quantify wind injection rates and durations using \texttt{MIST} tracks, identify the brief ($\sim10^4\,\text{yr}$) high-$\dot{M}$ phases of AGB mass loss, and show how concurrent wind-wind collisions in clusters with $M_{\text{GC}}\gtrsim 2\times10^5\,{\text{M}}_\odot$ lead to efficient gas trapping.
  \item In Section~\ref{sec:GasRet}, we compute collisionless retention fractions $f_{\text{ret}}(M,R_{\text{hm}})$ in Plummer model potentials and demonstrate that the resulting retention fractions naturally reproduce the observed increase in the second-population fraction $f_{\text{2P}}$ among Galactic GCs.
  \item In Section~\ref{sec:GasFate}, we examine the fate of the retained gas: dense gas cores ($\rho \gtrsim 10^{-18}\,\text{g}\,\text{cm}^{-3}$) are disrupted by stellar encounters on $t_{\text{enc}}\sim10^3$~--~$10^4\,\text{yr}$, precluding new star formation.  Instead, gas is accreted onto existing main-sequence stars and compact objects via Bondi-Hoyle accretion.
  \item In Section~\ref{sec:GasEvol}, we integrate wind injection and Bondi-Hoyle depletion into a core-halo gas mass evolution model, showing that a quasi-steady reservoir of $\lesssim 10^{-2}\,M_{\text{GC}}$ persists until the feedback turns on.
  \item In Section~\ref{sec:Consequences}, we discuss the implications for stellar population: how gas-driven accretion `rejuvenates' low-mass stars into the $4$~--~$6\,\text{M}_\odot$ range required for light element anomalies, and how compact object growth may initiate late feedback.
  \item We summarise our findings and their broader implications in Section~\ref{sec:Summary}.
\end{itemize}

In our second paper in the series (Paper II), we will quantify in detail how this gas-stellar interplay reshapes the compact object and main-sequence stellar populations.

\section{Stellar mass loss}

\label{sec:Winds}

In our study, we find that both the stellar distribution and stellar evolution have a significant influence on the amount of gas in globular clusters. In this section, we discuss the implications of stellar evolution on the presence of gas.

The total amount of mass released in stellar winds by a Kroupa population \citep{Kroupa2008} of stars over $10\,\text{Gyr}$ constitutes $20$~--~$25\,\%$ of their initial mass \citep{Thoul2002}. This is attributed to the fact that the mass of the white dwarf remnant, for those stars that do not explode as supernovae, is only a small fraction of the original stellar mass \citep{Kippenhahn1977}. Moreover, \citet{Thoul2002} found that more than half of the mass is released during the initial $1\,\text{Gyr}$ of the cluster's lifetime. This result can be attributed to the rapid evolution of the most massive stars, which release the largest amount of mass per star. Hence, particularly in the first $1\,\text{Gyr}$, the mass budget of the gas released into stellar winds is significant compared to the initial stellar mass.

\begin{figure}
    \centering
    \includegraphics{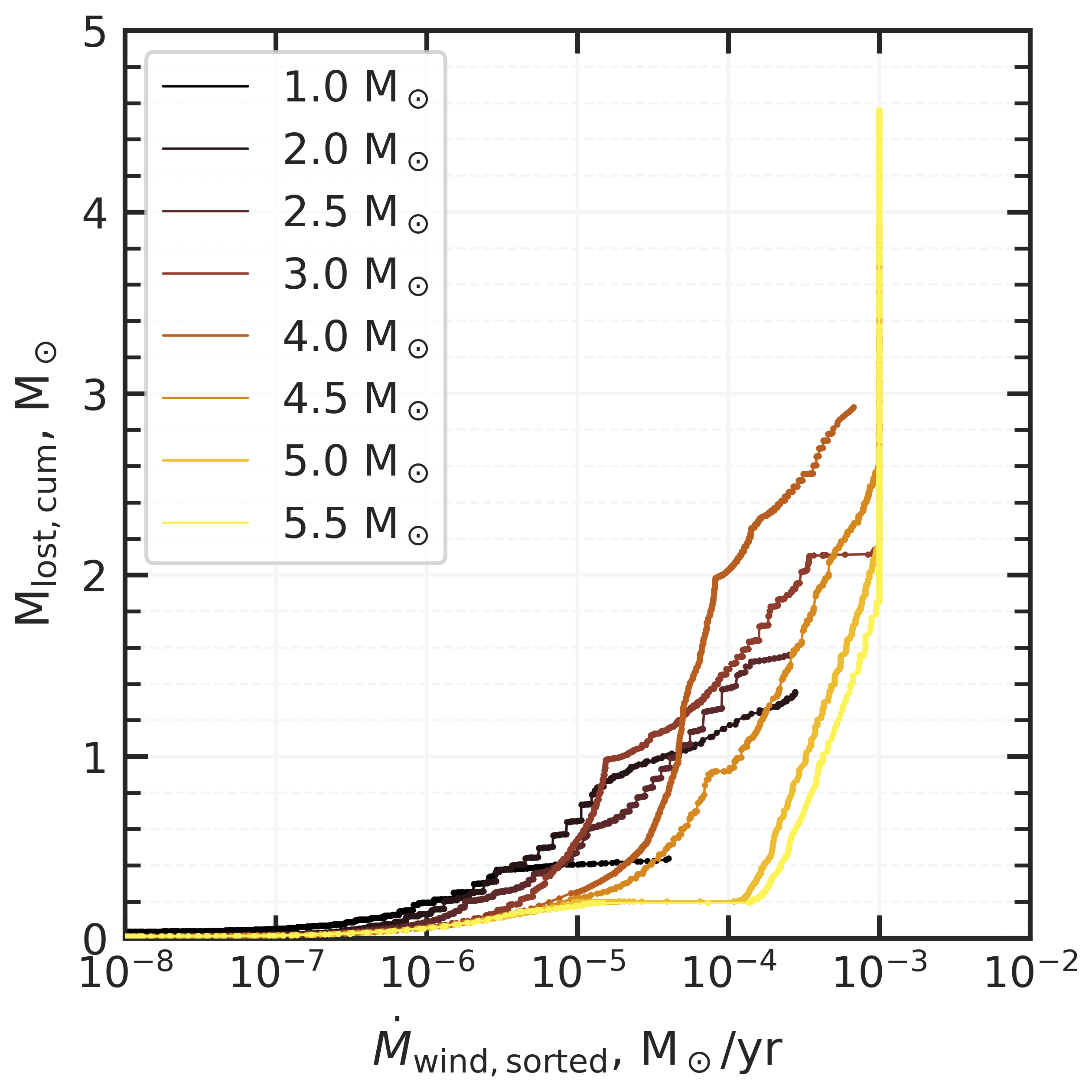}
    \caption{The cumulative amount of stellar mass lost through winds shown as a function of the wind mass loss rate $\dot{M}$, based on MIST stellar evolution tracks at $\text{[Fe/H]}=-1.5$. The cumulative mass loss is calculated by dividing the tracks into small time intervals and re-ordering these intervals in increasing order of $\dot{M}$. One can see that, according to the MIST stellar tracks, stars with initial masses above $2\,\text{M}_\odot$ lose most of their mass at rates higher than approximately $\dot{M}\gtrsim10^{-5}$~--~$10^{-4}\,\text{M}_\odot/\text{yr}$, corresponding to timescales of a few tens of kiloyears, which is relatively short compared to the final ages of the stars.}
    \label{fig:MDotSorted}
\end{figure}

Most of the material that stars lose through their winds is lost towards the end of their lifetimes during the thermally-pulsating AGB phase at high mass loss rates of $\dot{M}_{\text{wind}}\approx 10^{-4}$~--~$10^{-3}\,\text{M}_\odot/\text{yr}$. In Figure \ref{fig:MDotSorted}, we illustrate this by showing the cumulative amount of mass lost through winds as a function of the wind mass loss rate for several stellar masses. We construct the figure using MIST stellar evolution tracks \citep{Choi2016}, which are based on the detailed stellar evolution code MESA \citep{Paxton2011, Paxton2013, Paxton2015, Paxton2018, Paxton2019}. We divide the stellar tracks into small time intervals and sort them by wind mass loss rate to obtain the cumulative amount of stellar mass lost. As shown in the figure, all stars with initial masses above $2\,\text{M}_\odot$ finish their evolution within $1\,\text{Gyr}$, showing similar behaviour, losing the bulk of their material at comparably high wind mass loss rates. 

At mass loss rates of $\dot{M}_{\text{wind}}\approx 10^{-4}$~--~$10^{-3}\,\text{M}_\odot/\text{yr}$, during which most of the stellar mass is lost, the expected wind speed is approximately $v_{\text{wind}}\approx 20\,\text{km}/\text{s}$ \citep{Hoefner2018}. We will use $v_{\text{wind}}$ as a fiducial value, serving as a conservative upper estimate for the wind speed. This value is higher than the wind speeds observed in most AGB stars, which are in the range of $5$~--~$10\,\text{km}/\text{s}$. The difference arises because the wind speed of AGB stars increases with their mass loss rate \citep{Ramstedt2008}. Additionally, typical AGB stars spend only a small fraction of their lifetimes in such an active wind-losing phase and are hence rarely observed during such a phase. In summary, stars lose most of their mass at wind mass loss rates of $\dot{M}_{\text{wind}}\approx 10^{-4}$~--~$10^{-3}\,\text{M}_\odot/\text{yr}$, with corresponding wind speeds of $v_{\text{wind}}\approx 20\,\text{km}/\text{s}$.

\begin{figure}
    \centering
    \includegraphics{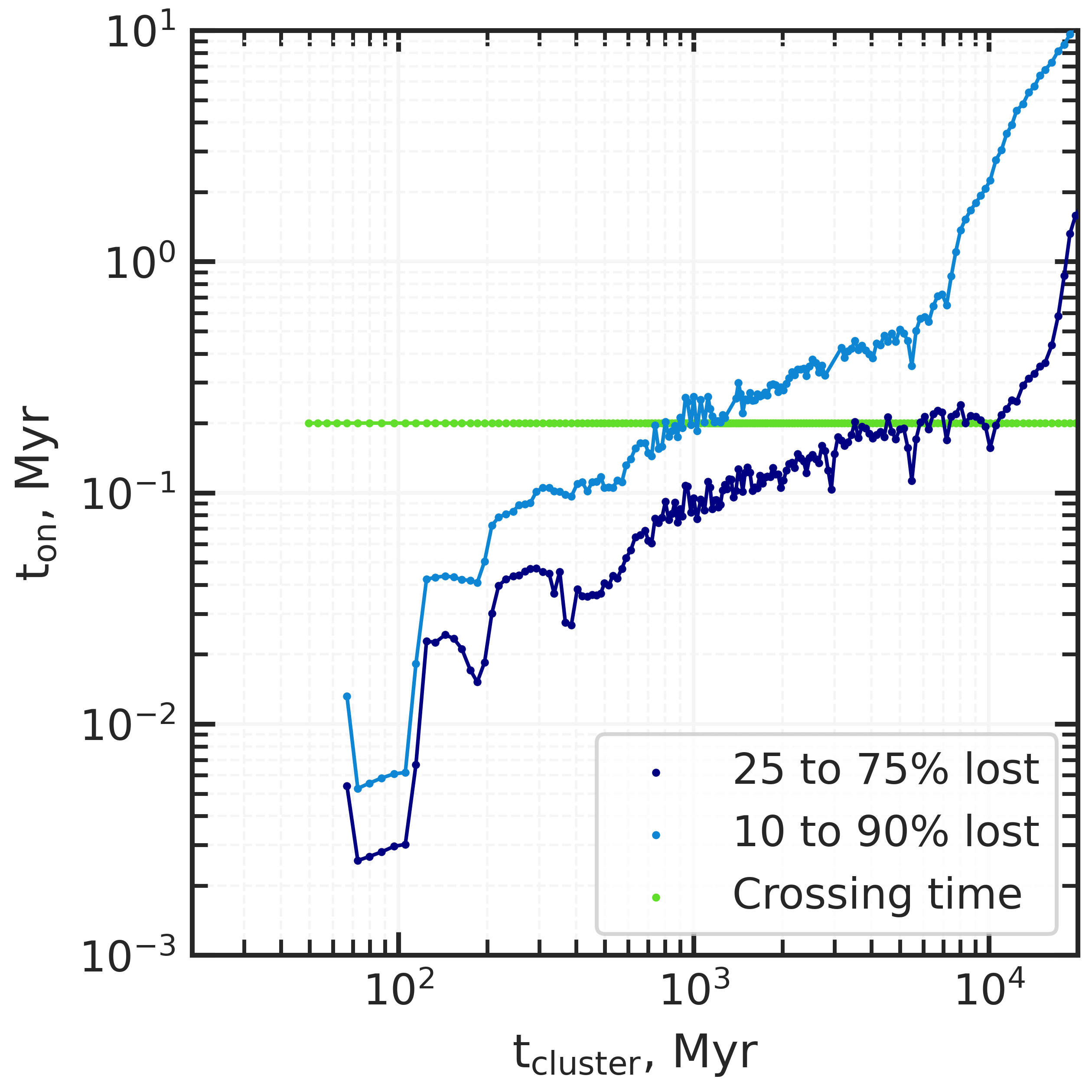}
    \caption{The duration of the active mass-losing phase for stars at $\text{[Fe/H]}=-1.5$ shown as a function of stellar lifetime. This phase is defined as the time interval between the points (in Figure~\ref{fig:MDotSorted}) when a star loses between $25\,\%$ and $75\,\%$ (dark blue line) or $10\,\%$ and $90\,\%$ (light blue line) of the total mass it is going to lose. Additionally, the green line in the plot shows the time required for the stellar winds to traverse the central $4\,\text{pc}$ region of the cluster. It can be concluded that, in the first $1\,\text{Gyr}$, the wind has a thick shell geometry relative to the cluster scale. In comparison, at later times, the wind outflow from a single star can engulf the entire cluster.}
    \label{fig:TOnTCluster}
\end{figure}

As a direct consequence of the wind mass loss history, the duration of the phase when most of the mass is lost is between a few $\text{kyr}$ and a few $10\,\text{kyr}$ for any stellar mass above $2\,\text{ M}_\odot$. In Figure~\ref{fig:TOnTCluster}, we show the duration of the wind mass loss phase for turnoff-mass stars as a function of cluster age. The duration of the wind mass loss phase is defined here as the time interval during which the star loses between $25$ and $75\,\%$ (or between $10$ and $90\,\%$) of its total lost mass, following the $\dot{M}$-ordered tracks. As shown in the figure, the duration of this active phase is systematically shorter early in the cluster's life, due to the more massive stars evolving more quickly. Moreover, during the first $\text{Gyr}$, the time it takes for the wind to traverse the central $4\,{\rm pc}$ of the cluster is typically several times longer than or comparable to the duration of this active wind mass loss phase. Therefore, during this epoch, the winds of each individual star have a thick shell geometry. Similarly, one may conclude that at later times, and especially at the current epoch, the winds of individual stars can engulf the entire cluster.

\begin{figure}
    \centering
    \includegraphics{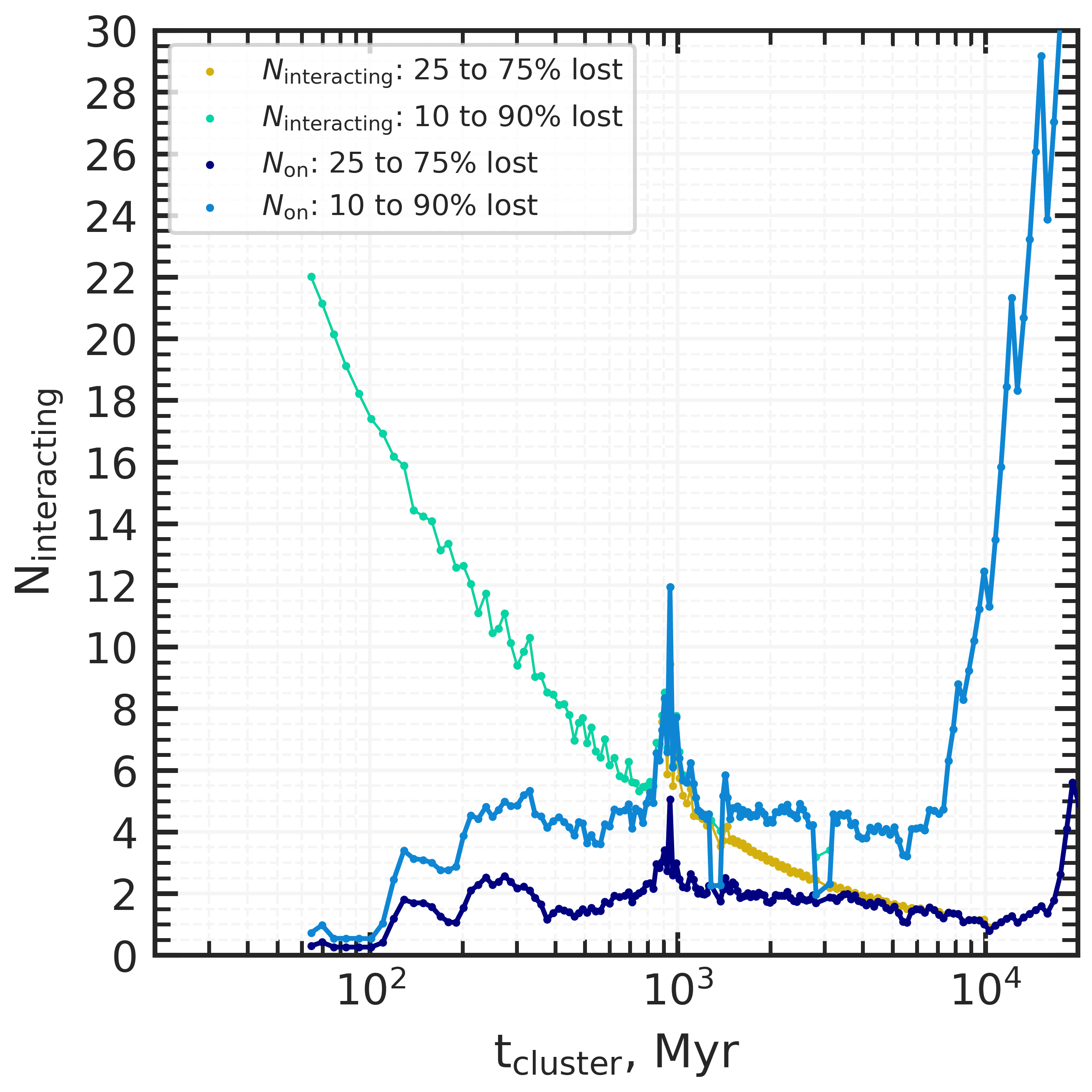}
    \caption{  
    The number of interacting stellar wind ejecta in a globular cluster with a Kroupa population of $N=10^6$ stars at a metallicity of $\text{[Fe/H]}=-1.5$, as a function of cluster age. The stellar winds of two stars are assumed to interact if either both stars are in the active mass-losing phase simultaneously or if the time for a wind ejecta shell to traverse the cluster is shorter than the time until the next star enters the active mass-losing phase. We can see that the number of interacting stellar wind ejecta in the globular cluster remains greater than one for most of the time until it reaches $10\,\text{Gyr}$. The collisions of stellar wind ejecta enhance gas retention in globular clusters.}
    \label{fig:NInt}
\end{figure}

In Figure~\ref{fig:NInt}, we show the total number of stars in the active mass-losing phase as a function of time for a Kroupa population of $N=10^6$ stars with a metallicity of $\text{[Fe/H]}=-1.5$, typical for a globular cluster. We calculate the number of stars in the `on' phase by determining the duration of this phase and the time interval between two consecutive stars completing their evolution, using the MIST tracks for interpolating the stellar ages. After $100\,\text{Myr}$, there are consistently between $3$ and $10$ stars in the active mass-losing phase. We can further infer that if multiple stars are `on' simultaneously, or if the time interval between two consecutive stars completing their evolution is shorter than the time required for the wind to traverse the cluster, their winds will collide within the cluster. To calculate the number of interacting stars in Figure~\ref{fig:NInt}, we use the condition that the shells collide either while traversing the cluster or due to simultaneous wind mass loss. The number of interacting shells is always at least several and can exceed $10$ during the first $\text{Gyr}$ of the population. Interacting shells lose kinetic energy through shocks, eventually radiating this energy and becoming more gravitationally bound to the cluster. Therefore, since the number of interacting shells is proportional to the number of stars $N$, clusters with at least $N_{\text{min}}\approx3\cdot 10^5$ stars will experience the interaction of stellar winds. As we show in the next section, this process contributes to wind retention in globular clusters.

\section{Wind retention in globular clusters}
\label{sec:GasRet}

\begin{figure}
    \centering
    \includegraphics{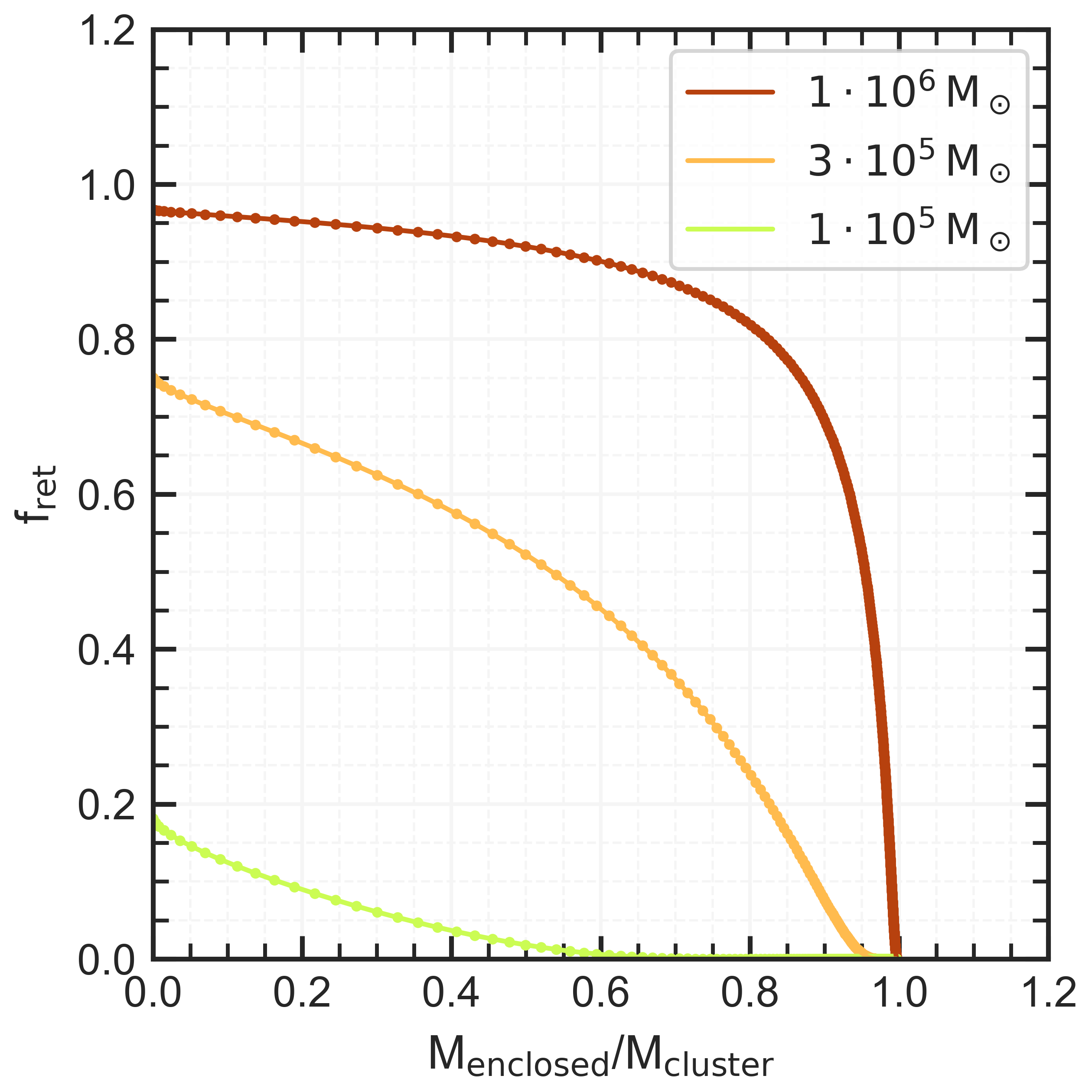}
    \caption{The fraction of stellar mass lost into winds that stays gravitationally bound to the host cluster as a function of enclosed mass within the cluster, normalised by the total mass. The figure is based on the Plummer model for $10^6\,{\rm M}_\odot$, $3 \times 10^5\,{\rm M}_\odot$ and $10^5\,{\rm M}_\odot$ clusters, shown by red, yellow and green lines, respectively. The cluster half mass radius is assumed to be $4\,{\text{pc}}$ and the wind speed is set to $20\,{\text{km}/\text{s}}$. More massive clusters of a given physical size have deeper potential wells, making it progressively easier for stellar winds to be retained within the cluster. At the cluster mass of $10^6\,{\text{M}}_\odot$, most of the stellar winds are gravitationally bound throughout the cluster.}
    \label{fig:RetentionFractions}
\end{figure}

\begin{figure}
    \centering    \includegraphics{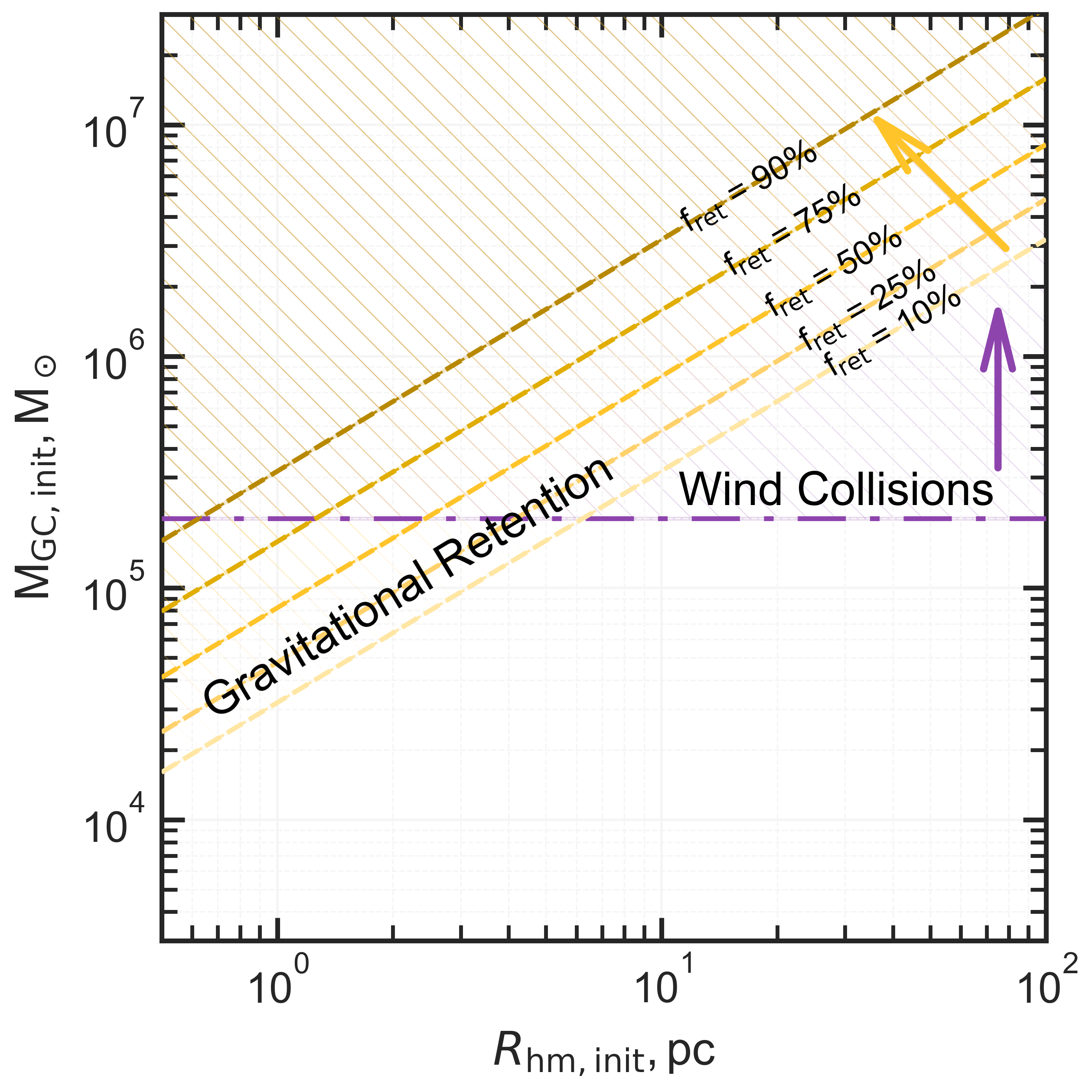}
    \caption{Gas retention fraction shown in the cluster initial mass -- half-mass radius plane. The yellow dotted lines are for gravitational gas retention fractions, $f_{\text{ret}}$, of $10\,\%$, $25\,\%$, $50\,\%$, $75\,\%$, and $90\,\%$ (from lower to upper lines) assuming a Plummer model for the cluster and all mass is lost via stellar winds having a speed of $20\,\text{km}/\text{s}$. The purple horizontal dot-dashed line represents the boundary above which multiple stars are likely to be undergoing high wind mass-loss rates at the same time, in which case collisions between winds will likely enable gas retention. Thus, gas will likely be retained in clusters above the horizontal dashed line or above the lowest of the dotted lines.}
    \label{fig:RetentionRegMHM}
\end{figure}

\begin{figure}
    \centering    \includegraphics{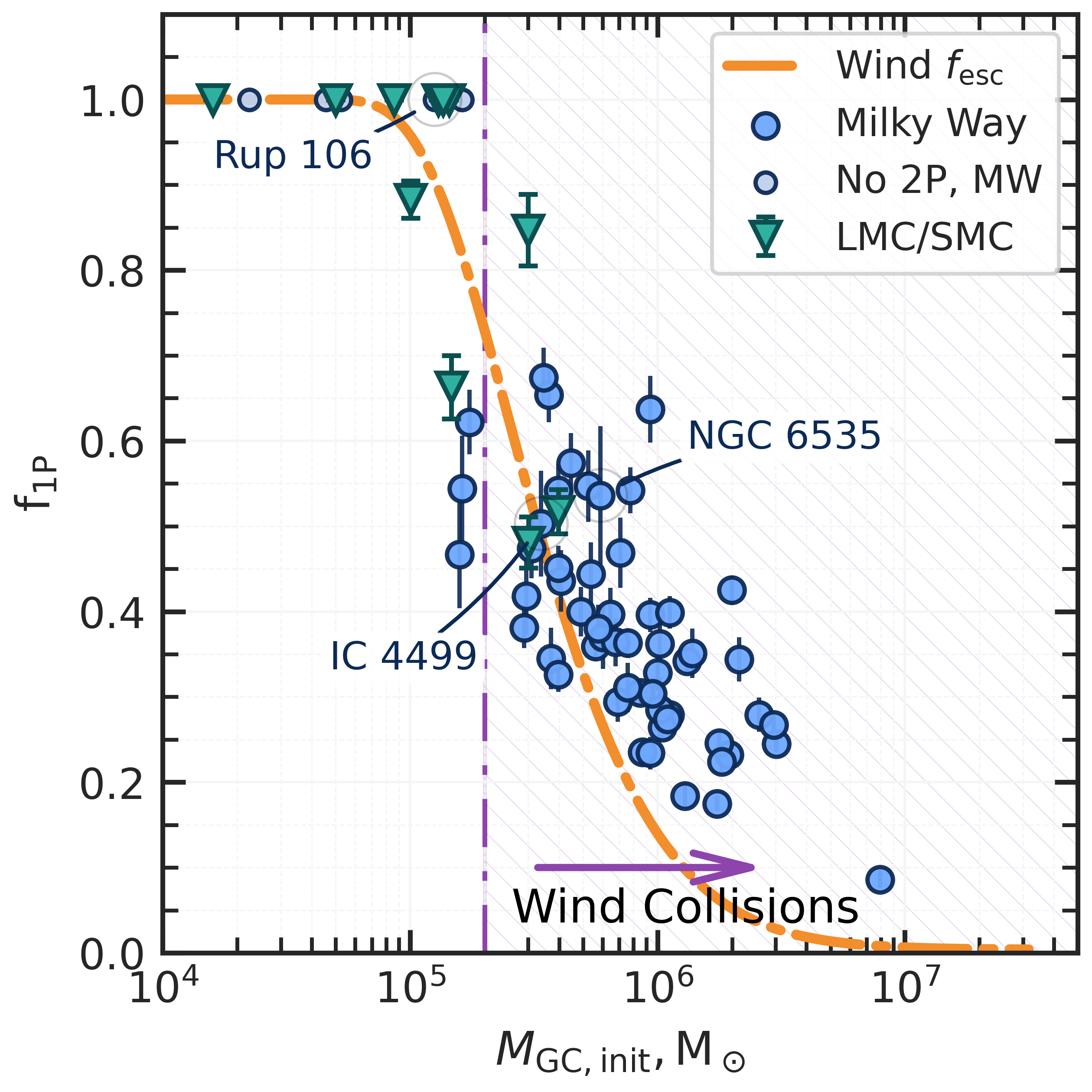}
    \caption{The fraction of stars which belong to the first population, $f_{\text{1P}}$, as a function of the initial cluster mass from the most recent version of the \citet{Baumgardt2018} catalogue derived from N-Body simulation-based fits to GC observations, for the sample of MW globular clusters given in \citet{Milone2017}, shown with blue markers, together with five MW globular clusters where little or no evidence of multiple populations has been found (left to right: Pal 12, E3, Ter 7, Rup 106, and Ter 8), shown with smaller light-blue markers. Additionally, using green triangles, we show the LMC/SMC clusters given in \citet{Milone20}, for which the most recent values for the initial masses were kindly provided by Holger Baumgardt (private communication). Shown with an orange dot-dashed line is the mass fraction of gas from stellar winds, which would escape from a Plummer model potential with a half-mass radius of $4\,\text{pc}$, assuming all mass is lost with a wind speed of $20\,\text{km}/\text{s}$. Additionally, we show the region where wind collisions are expected with a vertical purple dot-dashed line and a purple hatched region and show the names of the globular clusters discussed in the text. One can see that the clusters containing only one population are not expected to retain gas, according to the wind collision and gravitational gas retention mechanisms.}
    \label{fig:f_1P_vs_M_init}
\end{figure}

In this section, we investigate whether the winds released by stars at the end of their lifetimes can be retained in globular clusters. In our treatment here, we will initially consider stellar winds as a collisionless gas (i.e., neglecting hydrodynamical interactions) and determine whether the initial speed of the gas will be sufficient to unbind it from the potential well of the cluster.

We model stellar clusters as spherically symmetric Plummer models, with the density distribution given by \citet{Plummer1911, BinneyTremaine2008}:
\begin{equation} 
\rho (r) = \frac{3 M}{4 \pi a^3} \left(1 +  \frac{r^2}{a^2} \right)^{-5/2}.
\end{equation}
Here, $r$ is the radius, $M$ is the total cluster mass, and $a$ is a characteristic length scale. The gravitational potential is given by:
\begin{equation} 
\phi (r)  = - \frac{G M}{a} \left(1 +  \frac{r^2}{a^2} \right)^{-1/2}
\end{equation}
The distribution function, which represents both the locations and velocities of the stars, under the assumption that the velocities are locally isotropic, can be expressed as:
\begin{equation} 
f ({\bf r}, {\bf v}) {\rm d}^3 {\bf r} {\rm d}^3 {\bf v} = \dfrac{24 \sqrt{2}}{7\pi^3}\dfrac{a^2}{G^5 M^4}\left(-\phi(r) - \dfrac{v^2}{2}\right)^{7/2} {\text{d}}^3 {\bf r} \ {\text{d}}^3 {\bf v}
\label{eq:DF}
\end{equation}
for gravitationally bound stars satisfying $v\le v_{\rm esc}\equiv \sqrt{-2\phi(r)}$, and it is equal to zero otherwise.

Consequently, the three-dimensional velocity dispersion for the model is given by $\sigma_{\text{3D}}^2 (r) = - \phi(r)/2$. The radius $R_\alpha$, which encloses a mass $m_\alpha \equiv \alpha M$, is given by:
\begin{equation} 
R_{\alpha} = \frac{a}{\sqrt{\alpha^{-2/3}- 1}}
\label{eq:RAlpha}
\end{equation}
In particular, the half-mass radius,  $R_{\text{hm}}$, is given by  
\begin{equation} 
R_{\text{hm}}\equiv R_{0.5} = \frac{a}{\sqrt{0.5^{-2/3}- 1}} \simeq 1.305\,a
\end{equation}

One can consider the fate of gas from a star, moving at velocity ${\bf v}$, as a wind emitted isotropically at a wind speed $v_{\text{w}}$ in the rest frame of the star. If the combined velocity in a certain direction exceeds the local escape speed of the cluster, the gas may be considered lost. Conversely, if the total velocity is less than the local escape speed, the gas is gravitationally bound and, therefore, retained. For example, if for a given star $|{\bf v}| + v_{\text{w}} < v_{\text{esc}}(r)$, then the wind is gravitationally bound in all directions, and thus all the wind from such a star can be considered retained. Similarly, if for a certain star $|{\bf v}| - v_{\text{w}} > v_{\text{esc}}(r)$, then the wind is gravitationally unbound in all directions, and thus all the wind from such a star can be considered lost. For intermediate wind speeds, the retention fraction, $f_{\text{ret}}$, is calculated as follows:
\begin{equation}
    f_{\text{ret}} = \frac{1+\cos\theta}{2},
\end{equation}
while the wind escape fraction is $f_{\text{esc}} = 1-f_{\text{ret}}$. Here, $\theta$ represents the critical angle of the cone within which the wind is lost, and it is calculated as:
\begin{equation}
    \cos\theta = \frac{v_{\rm esc}^2 - v_{\text{w}}^2 - |{\bf v}|^2}{2 v_{\text{w}} |{\bf v}|}
\end{equation}

One can compute the total mass of gas from stellar winds retained within a cluster by integrating the distribution function in Equation~\ref{eq:DF} over all radii and velocities of the stars that are gravitationally bound to the cluster. The result of such a calculation is shown in Figure~\ref{fig:RetentionFractions}, where we plot the fraction of stellar mass lost into winds that remain gravitationally bound to the host cluster, $f_{\text{ret}}$, as a function of the enclosed mass within the cluster, normalised by the total mass. The figure is based on the isolated Plummer model described above, having a half-mass radius $R_{\text{HM}} = 4\,{\text{pc}}$, and for total cluster masses of $10^6\,{\text{M}}_\odot$, $3 \times 10^5\,{\text{M}}_\odot$, and $10^5\,{\text{M}}_\odot$, which are shown by red, yellow, and green lines, respectively. We set the wind speed to $20\,{\text{km/s}}$ for all the stars following \citet{Hoefner2018}, as discussed in Section~\ref{sec:Winds}.

We can see from Figure~\ref{fig:RetentionFractions} that clusters with masses less than $10^5\,{\text{M}}_\odot$ can retain only a small fraction of the gas. In contrast, clusters with masses greater than $10^6\,{\text{M}}_\odot$ have sufficiently deep gravitational potential wells to retain most of the gas. Wind retention effectively `turns on' at this range of globular cluster masses because the cluster escape speeds are comparable to the stellar wind speeds. 

Two additional effects will enhance gas retention within globular clusters: the collisions between stellar winds and the interaction between winds and any blanket of previously retained gas.

Collisions between stellar winds will occur when multiple stars experience high rates of mass loss simultaneously. Gas retention becomes more likely in such situations, as collisions between stellar winds convert the outgoing kinetic energy into thermal energy, which can then be radiated away. Therefore, gas retention is likely to be enhanced in clusters that are massive enough to host multiple stars that emit significant winds simultaneously.

The process of wind collisions is primarily governed by the number of stars in the population and by the stellar evolution processes, as discussed in Section~\ref{sec:Winds}. As seen in Figure~\ref{fig:TOnTCluster}, the phase of high mass loss rates in stars is relatively short, particularly during the first $\text{Gyr}$, ranging approximately between $10\,{\text{kyr}}$ and $100\,{\text{kyr}}$ during that time. Consequently, as Figure~\ref{fig:NInt} shows, for a cluster initially containing $10^6$ stars, the number of stars in this phase during the first $\text{Gyr}$ is likely to be a few. Additionally, stellar winds typically take about one crossing time to leave the cluster. As Figure~\ref{fig:NInt} also indicates, the number of simultaneously traversing stellar wind shells in such a cluster is roughly between $5$ and $15$. Therefore, one can generally expect about one stellar wind shell traversing the cluster per $10^5\,{\rm M}_\odot$ of stars, requiring a minimum cluster mass of $M_{\rm cluster} = 2 \times 10^5\,{\rm M}_\odot$ for wind-wind collisions. Here, we reiterate that stellar wind collision processes, which depend on the details of late stellar evolution, tend to become significant at cluster masses similar to those needed for gravitational potential wind retention.

Finally, as previously noted, gas retention itself further promotes more gas retention. In other words, if initially some gas is retained, this gas forms a blanket that enables the retention of subsequent winds. Given the extremely short phase of significant mass loss (as shown in Figure~\ref{fig:TOnTCluster}), the amount of gas previously accumulated in the blanket rapidly exceeds the mass each star produces, effectively absorbing the outgoing momentum. Therefore, gas retention is expected to function as a switch: if gas retention is possible, a significant portion of the gas is likely to be retained in a cluster.

Based on the considerations above, we can use the initial mass -- half-mass radius plane of globular clusters to indicate the two regions where gas retention is likely to occur: clusters with a sufficiently deep gravitational potential well, and also those with a sufficiently large mass, independent of cluster radius. We plot these regions in Figure~\ref{fig:RetentionRegMHM}. The yellow dashed lines show gas retention fractions, $f_{\text{ret}}$, of $10\,\%$, $25\,\%$, $50\,\%$, $75\,\%$, and $90\,\%$ (from lower to upper lines), assuming a Plummer model for the cluster and that all mass is lost via stellar winds with a speed of $20\,\text{km}/\text{s}$. The horizontal purple dot-dashed line represents the boundary above which multiple stars are likely to be undergoing high mass loss rates at the same time. The regions in which either or both mechanisms operate indicate three potentially distinct regimes for gas retention in clusters. Finally, it is worth noting that gas retention, as will become evident in the next paragraph, acts as a relatively smooth switch. Therefore, the lines in Figure~\ref{fig:RetentionRegMHM} should be seen as indicative of the transition from a gas-free regime to a gas-retaining regime rather than sharp boundaries. These results are also consistent with the more simplified analysis in \cite{Mas+21}, which showed that the second population fraction correlates with globular cluster escape speeds for typical AGB wind velocities.

It is important to recall the distinction between initial and current cluster masses, as clusters may lose a significant fraction of their mass over time due to internal processes and tidal stripping. Since gas retention is most important in the first Gyr of globular cluster history, the initial masses are more relevant for the observed correlation. Following this reasoning, we used the values for the initial masses from Version~4 of the online Baumgardt catalogue of globular cluster properties, which is based on N-Body simulations and detailed photometric fits to present-day clusters \citep{Baumgardt2017, Baumgardt2018, Sollima2019, Baumgardt2020, Baumgardt2021, Baumgardt2023}. For example, it is informative to compare the two Galactic globular clusters Rup~106 and NGC~6535, highlighted in Figure~\ref{fig:f_1P_vs_M_init}. Observations reveal that Rup 106 is notably unusual as it is a globular cluster that does not contain multiple stellar populations \citep{Dotter2018}, whereas NGC 6535 possesses multiple populations, with $f_{\text{1P}} = 0.536$ \citep{Milone2017}. According to the online Baumgardt catalogue, the current masses for these two clusters are $3.4 \times 10^4\,{\text{M}}_\odot$ and $2 \times 10^4\,{\text{M}}_\odot$, respectively. However, when comparing the initial masses, it is evident that Rup~106 was the less massive of the two clusters, having its initial mass around $1.26 \times 10^5\,{\text{M}}_\odot$ compared to $5.89 \times 10^5\,{\text{M}}_\odot$ for NGC~6535. Therefore, initially, Rup~106 had a significantly lower capacity for retaining gas compared to NGC~6535. Similarly, globular cluster IC~4499, also highlighted in Figure~\ref{fig:f_1P_vs_M_init}, currently having mass of $1.50\times 10^5\,{\text{M}_\odot}$, had its initial mass equal to approximately $3.39\times 10^5\,{\text{M}_\odot}$. Therefore, the initial and present-day masses of IC~4499 resemble those of NGC~6535 rather than Rup~106, consistent with the multiple populations reported in IC~4499 \citep{Dalessandro2018}.

Observations of young clusters (ages up to $300\,{\text{Myr}}$) in the Magellanic Clouds have placed significant limits on the amount of gas present within them \citep{Bastian2014}. The authors concluded that the clusters they studied could contain at most $75\,{\text{M}_\odot}$ of gas, as indicated by dust measurements, or at most $200\,{\text{M}_\odot}$ of gas, as indicated by HI measurements, typically amounting to $<1\,\%$ of the total cluster mass. In other words, observations of young clusters in the Magellanic Clouds would seem to imply that gas from stellar winds is not retained. We show the globular clusters in the LMC and SMC in Figure~\ref{fig:f_1P_vs_M_init}, with the second population fractions obtained from \citet{Milone20} and the updated most recent values of their initial masses (compared to \citealt{Milone20}) kindly provided by Holger Baumgardt (private communication). One can see that when considering the initial cluster masses, one does not expect these young clusters to retain significant amounts of gas and hence form multiple populations. More broadly, one should consider the depth of the gravitational well in the clusters since the LMC/SMC globular clusters are also slightly larger compared to the Galactic GCs, and have half-mass radii of approximately $8\,{\text{pc}}$. Therefore, based on our calculations presented here, we would also expect the clusters observed in \citet{Bastian2014} to be devoid of gas as their gravitational potential wells would be too shallow to retain the gas released as stellar winds.

In Paper II of this series, we will explore in detail the idea that gas retained in sufficiently massive or compact globular clusters leads to the production of the second population of stars seen in some globular clusters \citep[e.g.,][]{Conroy2011}. Here we simply note that {\it the presence (and indeed size) of a second stellar population appears to correlate with the retention of gas from winds derived from the first population of stars}, as shown in Figure~\ref{fig:f_1P_vs_M_init}.

\section{Fate of retained gas}
\label{sec:GasFate}

In this section, we consider the fate of the retained gas. It has been suggested that such gas could directly form the second population of stars observed in some globular clusters \citep[e.g.][]{DErcole2008,Conroy2011}. However, below, we show that encounters between pre-existing first-population stars and gaseous protostellar cores (which would otherwise form stars if left alone) occur on sufficiently short timescales to inhibit the formation of new stars. Instead, we find the gas is more likely to be accreted onto a subset of the pre-existing stars. Most likely, these stars will be found towards the cluster centre, as retained gas is likely to settle in the central regions. As shown below, several feedback mechanisms could ultimately remove essentially all of the gas. As described below, such mechanisms are likely to turn on over time, perhaps as late as around $1000\,\text{Myr}$, leaving all present-day globular clusters virtually devoid of gas, thus matching observations. We note, however, that the feedback timeline is a significant source of uncertainty in our model, and detailed studies of the feedback sources and their progenitors in gaseous and dynamical environments are necessary to accurately constrain these processes.

\subsection{Star formation is delayed due to ionising radiation}

The first population of globular cluster stars can lose up to about 10\,\% of the total stellar mass into winds within the first few $100\,\text{Myr}$ \citep{Thoul2002}. Given the high total mass of retained gas, it is therefore likely to reach densities of order $\gtrsim 10^4\,\text{M}_\odot/\text{pc}^3\approx 7\times 10^{-19}\,\text{g}/\text{cm}^3$ comparable to those found in the dense cores of molecular clouds, e.g.\,\citet{Imara2023}. Therefore, it would be reasonable to assume that such gas would cool and fragment to form denser cores and thence further contract to form stars. Indeed, a common feature of many models for the production of a second stellar population is that the second population forms directly from the gas, possibly mixed with some un-enriched gas \citep[see, for example,][]{DErcole2008}.

However, the formation of stellar cores is expected to be delayed because of a lack of molecular hydrogen, which is a critical building material. As \citet{Conroy2011} pointed out, young stellar populations produce a significant amount of ultraviolet radiation that is capable of ionising atomic and molecular hydrogen, thus inhibiting star formation. Within a few $100\,\text{Myr}$, as the turnoff stars become cooler, the ionising flux drops by several orders of magnitude, allowing molecular hydrogen to form on dust grains and triggering a second episode of star formation.

\subsection{Star formation is inhibited by encounters}
\label{sec:StarForm}

We consider below whether the star formation process will be interrupted due to encounters between cores and pre-existing stars. Protostellar cores have sizes around $10^4\,\text{AU}$ \citep[see, for example,][]{McKee2003}. At such separations, the relative circular Keplerian velocity of two stars is much smaller than the velocity dispersions within the massive stellar clusters considered here. Therefore, gravitational focusing will not play a role in encounters having such large pericentre separations, and we can use a simple geometric approximation for the physical collision cross-section. Thus, for stellar density $n_\star$, protostellar core radius $r_{\text{lump}}$ and relative velocity $v_\infty$, the encounter timescale is given by
\begin{equation}
t_{\text{enc}} \simeq \frac{1}{n_\star \pi r_{\text{lump}}^2 v_\infty}
\end{equation}
Furthermore, assuming $r_{\rm lump} = \beta\,\text{AU}$, $v_\infty = x\,\text{km}/\text{s}$,  $n_{\star,5}  = n_\star/(10^5\,\text{stars}/\text{pc}^3)$, we can rewrite the encounter timescale as 
\begin{equation}
t_{\text{enc}} \simeq \frac{4 \times 10^5}{n_{\star,5} \pi \beta^2 x}\,\text{Myr}.
\end{equation}
Considering conditions relevant for the core of a typical enriched globular cluster, $x=30$, $n_{\star,5}  =  1$, and taking $\beta=10^4$, we obtain $t_{\text{enc}}\simeq 40\,\text{yr}$. In comparison, taking conditions more applicable to the half-mass radius of a cluster, i.e.\, $x=10$, $n_{\star,5}  =  10^{-2}$ and still using $\beta=10^4$, we obtain $t_{\text{enc}} \simeq 1.3 \times 10^4\,\text{yr}$.

This timescale should be compared to the timescale for the evolution of the cores as the gas contracts to form stars. This occurs on a timescale of $\sim 10^5\,\text{yr}$ \citep{McKee2003}. Therefore, we see that {\it direct star formation within the retained gas is likely to be disrupted via encounters with pre-existing stars}. Indeed, \citet{Bekki2019} noted that star formation is inhibited by a stellar population embedded in gas. A similar effect has also been seen in hydrodynamic simulations of star formation, where pre-existing stars tidally break up other cores before they can form stars, e.g.,~\citet{Smilgys2017}.

Therefore, rather than directly form a second population of stars, we believe this gas will be accreted onto pre-existing stars, including any compact objects which may be found within the cluster. The subsequent evolution of compact objects and main-sequence stars due to accretion will be considered in Sections~\ref{sec:GasEvol}~and~\ref{sec:Consequences}. 

\subsection{Gas cloud geometry}
The retained gas carries significant angular momentum relative to the central stars, as was, e.g., modelled by \citet{Bekki10} and \citet{MastrobuonoBattisti13}. Since retained gas comes from the whole cluster, it must contract as it enters the central regions and consequently acquire angular velocity relative to the central stars. 

Typical globular cluster rotation speeds $v_{\text{rot},\star}$ are usually of order a few $\text{km}/\text{s}$ but in some cases may reach $10\,\text{km}/\text{s}$ \citep{Bellazzini12,Baumgardt2018}. Stellar wind from typical stars at the half-mass radius, after reaching the area with an outer radius $R_{\text{gas}}$, will have a net rotational velocity of order $v_{\text{rot,gas}}\approx (2
^{1/3}R_{\text{hm}}/R_{\text{gas}}) v_{\text{rot},\star} \approx 2 v_{\text{rot},\star}$ (assuming $R_{\text{gas}}$ corresponds to the quarter-mass radius of the cluster). Therefore, gas will have a non-negligible rotation speed, typically reaching about $5\,\text{km}/\text{s}$ and in some cases as much as $20\,\text{km}/\text{s}$, likely taking the shape of an ellipsoid and having net rotation relative to central stars.

Furthermore, the gas cloud may also flatten significantly, depending on its cooling efficiency, as discussed in \citet{Mas+16,Perets2022}. For the environment of B-type stars (which are more massive than about $2\,\text{M}_\odot$), the typical ambient temperature may be several $10^4\,\text{K}$, which corresponds to sound speeds $5$~--~$10\,\text{km}/\text{s}$. Therefore, a somewhat flattened disc-like configuration defined by $h/R\approx \sigma_v/c_s \sim 1$ is plausible. 

The gas structure will also be continually disturbed by the embedded stars through gas-dynamical friction, gas accretion by stars, and stellar feedback, e.g., \citet{Grudic22}. The cooling gas may also form filamentary structures. Therefore, a realistic distribution of gas in a globular cluster environment will be both relatively complicated and tend to exhibit central concentration and net rotation.

\subsection{Viability of Bondi-Hoyle accretion}

We consider here Bondi-Hoyle (B-H) accretion of gas directly onto stars. We note, however, that the model of B-H accretion contains uncertainties due to its inherently complex nature. Specifically, during B-H accretion, a star produces a convergent flow of ambient gas that loses energy due to self-intersections and subsequently accretes onto the star through a disc. The convergent flow depends on whether the stars move subsonically or supersonically relative to the ambient gas, and the accretion feedback may also modify the flow \citep{Edgar2004}. A sufficiently massive accretion disc may also become unstable to forming self-gravitating lumps through the Toomre instability. It has been suggested by \citet{Winter2023} that massive planets/low-mass stars form within such discs, which are then subsequently scattered into their hosting stars on a timescale of a few Gyr, thus polluting stars but only at a later time.

For the calculations in this paper, we use the standard expression for the Bondi-Hoyle accretion given by 
\citet{Edgar2004}:
\begin{equation}
\dot{M}_{\text{B-H}} = \dfrac{4\pi \rho G^2 M^2_\star}{(c_{\text{s}}^2 + v^2)^{3/2}}
\label{eq:BH1}
\end{equation}
\[
= 3.51\times 10^{-9} \left(\dfrac{M_\star}{\text{M}_
\odot}\right)^2 \left(\dfrac{\rho}{10^{-18}\,\text{g}/\text{cm}^3}\right)\left(\dfrac{10\,{\text{km}/\text{s}}}{c_{\text{s}}^2 + v^2}\right)^{3/2}\,\dfrac{\text{M}_\odot}{\text{yr}}
\]
Here, $M_\star$ is the mass of the star moving with speed $v$ relative to the background gas with density $\rho$ and sound speed $c_{\text{s}}$.



One can gain some insight into the effects of Bondi-Hoyle growth on stars by considering an individual star moving in a background gas, assuming $\rho$, $c_{\text{s}}$, and $v$ have constant values. Under such conditions, the growth rate of a given star, $\dot{M}_\star = \dot{M}_{\text{B-H}} = K M_\star^2$. This may be solved analytically, to give: 
\begin{equation}
M_{\star}(t) = \frac{M_{\star,\text{init}}}{1 - \left( t - t_{\text{init}} \right) K M_{\star,\text{init}}},
\label{eq:BH2}
\end{equation}
where $M_{\star,\text{init}} \equiv M_{\star}(t_{\text{init}})$ is the initial mass of the star and $K \equiv {4\pi \rho G^2}{(c_{\text{s}}^2 + v^2)^{-3/2}}$ is a constant. As can be seen from the above equation, the stellar mass increases to much larger values once $(t - t_{\text{init}})K M_{\star,\text{init}} \sim 1$. Thus, significant stellar-mass growth will occur on a characteristic timescale given by (see also \citealt{Leigh2013,Davies2020}):
\begin{equation}
\tau_{\text{B-H}}\equiv \dfrac{M_{\star,\text{init}}}{\dot{M}_{\text{B-H}}} = \dfrac{1}{K M_{\star,\text{init}}}
\end{equation}

Consider a gas reservoir with a total mass of $M_{\text{gas}} = 10^4\,\text{M}_\odot$, which is about 1 per cent of the total mass of clusters considered in this paper, contained in a characteristic volume of $1\,\text{pc}^3$, a volume which typically contains about 25 per cent of all stars in a cluster. In this case, $\rho = 6.8\times 10^{-19}\,{\text{g} \   \text{cm}^{-3}}$. Using Equation~\ref{eq:BH1}, we find $\dot{M}_{\text{B-H}} = 2.4\times 10^{-9}\,\text{M}_\odot/\text{yr}$ for $M_\star = 1\,\text{M}_\odot$ and $\sqrt{v^2 + c_\text{s}^2} = 10\,\text{km}/\text{s}$. These conditions yield a  characteristic timescale, $\tau_{\text{B-H}} \sim 400\,\text{Myr}$. In other words, {\it reasonable conditions for the gas cloud can lead to significant growth of stellar masses on relevant timescales}.


The realistic evolution of a given star is governed by the interplay of its own growth through Bondi-Hoyle accretion, its stellar evolution timescale, and the timescale of the depletion of the gas reservoir by the entire stellar population. Assuming a constant gas reservoir density, we can estimate the mass that stars can gain through Bondi-Hoyle accretion before they leave the main sequence. Indeed, the main-sequence lifetime of a $4.5\,\text{M}_\odot$ star, $\tau_{\text{MS},4.5}\approx 100\,\text{Myr}$, equals the Bondi-Hoyle growth timescale of $2\,\text{M}_\odot$ stars for the gas-cloud conditions considered above.  Hence, stars can grow to masses such that stellar winds are enriching, i.e. $M_\star\sim 4.5 - 6.8\,\text{M}_\odot$. Bondi-Hoyle growth from winds could, therefore,  {\it increase the population} of enriching stars, which in turn increases enrichment over time, thus helping to produce the observed enriched second population in globular clusters. Using detailed simulations in Section~\ref{sec:GasEvol}, we will show that gas density evolution in globular clusters supports this scenario.

\subsection{Feedback mechanisms}
\label{sec:Feedback}

The gas-rich epoch in the life of globular clusters eventually comes to an end, as demonstrated by the significant gas depletion observed in their present-day Galactic population. As an example, the gas content in Galactic globular clusters is considerably smaller than the amount expected from the winds of individual stars \citep{Chantereau2020}. In this section, we review the models for several feedback processes that have been proposed to be responsible for the observed gas depletion, highlighting the typical uncertainties associated with each of these processes.

Gas ram pressure stripping occurs every few $100\,\text{Myr}$ when GCs pass through the Galactic midplane and can presently remove between $10^2$ and $10^3\,\text{M}_\odot$ of gas per passage \citep[see, for example,][]{Tayler1975, Roberts1988}. In the early Universe, during the gas-rich epoch of the Galactic GCs, ram pressure stripping was likely even more effective due to the much higher gas content in the Galactic disc \citep{Conroy2011,Agertz2021}. Thus, midplane crossings could potentially remove all or a significant portion of GC gas. However, even complete gas removal cannot end the gas-rich epoch in GCs because the winds of AGB stars rapidly replenish the intracluster gas reservoir. Gas returns to equilibrium between wind injection and B-H accretion on a timescale that can be estimated as the time it takes for stellar winds to refill the amount of gas leading to interesting B-H accretion, $0.01M_{\text{GC}}/(0.1M_{\text{GC}}/100\,\text{Myr})\approx 10\,\text{Myr}$, which we also confirm using detailed models in Section~\ref{sec:GasEvol}. In summary, the period to restore the equilibrium gas amount is short compared to the interval between periastron passages.

Bondi-Hoyle accretion of gas onto compact objects, such as white dwarfs (WDs), neutron stars (NSs), or black holes (BHs), may continually unbind the GC gas, provided the energy deposition is not too anisotropic.

Consider the gravitational binding energy of the gas of mass $M_{\text{gas}}$ with the host GC, which has a central escape velocity $v_{\text{esc}}$:
\begin{equation}
E_{\rm bind}=\dfrac{M_{\text{gas}} v_{\text{esc}}^2}{2}=3.6\times 10^{50}\,\text{erg}\left(\dfrac{M_{\text{gas}}}{10^4\,{\text{M}}_\odot}\right)\left(\dfrac{v_{\text{esc}}}{60\,{\text{km}/\text{s}}}\right)^2
\label{eq:GasBind}
\end{equation}

At the same time, the accretion luminosity onto individual compact objects may be expressed as:
\begin{equation}
    L_{\text{acc}} = \eta \dot{M}_{\text{B-H}}c^2 = 5.7\times 10^{36}\dfrac{\eta}{0.1}\dfrac{\dot{M}_{\text{B-H}}}{10^{-9}\,\text{M}_\odot/\text{yr}}\,\dfrac{\text{erg}}{\text{s}},
\label{eq:AccLum}
\end{equation}
where the accretion efficiency $\eta$ is typically $0.1$ for BH accretors, $0.2$ for NS accretors, and $10^{-4}$ for WD accretors \citep{Frank2002}. Therefore, using Equation~\ref{eq:BH1}, for gas mass $M_{\text{gas}}=10^4\,\text{M}_\odot$ contained in a characteristic radius of $1\,{\text{pc}}$, in a GC with escape speed $v_{\text{esc}} = 60\,\text{km}/\text{s}$ and for $\sqrt{v^2 + c_\text{s}^2} = 10\,\text{km}/\text{s}$, a black hole of mass $10\,{\text{M}}_\odot$ will accrete gas at a rate of $\dot{M}=2.4\times 10^{-7}\,{\text{M}}_\odot/{\text{yr}}$. Then, using Equations~\ref{eq:GasBind} and \ref{eq:AccLum}, Bondi-Hoyle accretion onto a single BH can potentially deposit sufficient energy into the gas to unbind all of it in about $10\,{\text{kyr}}$, assuming maximally effective energy deposition. For a single NS accretor, due to its lower mass compared to BH accretors, the gas removal timescale is longer but still short, being a few $100\,{\text{kyr}}$. In contrast, single WDs have a gas removal timescale of the order of $100\,{\text{Myr}}$ due to their lower mass and much lower accretion efficiency (and accretion rate). However, a reasonable WD population will reduce the timescale to less than $1\,\text{Myr}$.



In realistic scenarios, feedback from B-H accretion may be delayed by the birth kicks received by the BH and NS accretors. These kicks can be sufficiently large to remove many BHs and NSs from globular clusters. Those that do remain bound to their birth cluster are expected to be located initially at larger radii and certainly outside of any gas cloud forming within the central core of a cluster. The heavier BHs and NSs will, however, sink back to the core via dynamical friction within a few  $100\,\text{Myr}$ \citep[for example, ][]{DaviesBenz1995, Morscher2015}. Furthermore, feedback from individual black holes may be significantly beamed, which would imply that hundreds of compact objects may be needed to affect the gas reservoir. In addition, the feedback process from a given BH or NS (or a BH subcluster) may self-regulate B-H accretion onto them, rather than depositing energy into the whole reservoir, thus reducing the accretion rate and the overall feedback. The onset of feedback via accretion onto BHs and NSs may therefore be effectively delayed, although detailed simulations are required to understand the relevant timescales over which this feedback channel becomes active.

Another important aspect to remember is that only some of the accretion luminosity will be absorbed by the gas. Firstly, gas may be sufficiently optically thin to allow a fraction of the radiation to escape the cluster. The optical thickness of the gas cloud, from its centre to its outer radius, can be estimated by assuming Thomson scattering, representing the upper limit of gas opacity:
\begin{equation}
    \tau = \dfrac{4}{9}\dfrac{\pi \sigma_{\text{T}}}{m_{\text{p}}} \dfrac{\bar{Z}}{\bar{A}}\dfrac{M_{\text{gas}}}{R_{\text{gas}}^2} = 0.83\left(\dfrac{M_{\text{gas}}}{10^4\,\text{M}_\odot}\right)\left(\dfrac{R_{\text{gas}}}{1\,\text{pc}}\right)^{-2}\left(\dfrac{\bar{Z}/\bar{A}}{5/7}\right),
\end{equation}
where $m_{\text{p}}$ is the proton mass, $R_{\text{gas}}$ is the radial density scale height, and $\sigma_{\text{T}}=6.65\cdot10^{-33}\,\text{cm}^2$ is the Thomson cross-section, and a pure hydrogen/helium mixture approximation is assumed. We can see that a decrease in gas mass or an increase in its radius can regulate accretion feedback. If the gas has lower opacity due to non-ionised atoms, if the accretion feedback is significantly beamed, or if the gas cools down before being ejected, the feedback efficiency may be substantially reduced further. Under such conditions, accretion onto many BHs or NSs (rather than a single object) may be required to unbind the gas.

It is also conceivable that an intermediate mass black hole (IMBH) forms in a globular cluster \citep{Baumgardt2019,Haberle24}. As discussed in \citet{Leigh2013}, such an IMBH may dominate B-H accretion and rapidly deplete the gas reservoir either by accretion or stellar feedback. If such an IMBH forms early on, e.g. through runaway stellar collisions, it may quench the gas-rich epoch early on (unless accretion physics prevents the process) and significantly reduce the effects of gas on stellar populations. Alternatively, as discussed in \citet{Leigh2013}, if one or more such IMBHs (or very massive stellar black holes) form much later as a result of gas accretion on stellar-mass BHs, they may deplete the gas and hence their formation may mark the end of the gas-rich phase. 

We finally consider feedback mechanisms related to binary stellar evolution, many of which will be affected by the dynamical interactions and gaseous environments that will occur within clusters.

Stars more massive than around $7\,\text{M}_\odot$ will explode as core-collapse supernovae (CCSNe). These explosions are likely to have sufficient energy to unbind gas within a cluster. Lifetimes of stars producing CCSNe range between about $3$ and $60\,\text{Myr}$ \citep{Choi2016}. Thus, we would expect any gas remaining from the star formation phase to be removed from the cluster at this time. Indeed, it is the gas released in stellar winds later, when lower-mass stars evolve, which we consider in this paper.

However, stars sufficiently massive to produce CCSNe can be produced as the outcome of binary evolution. For example, if two $5\,\text{M}_\odot$ stars in a binary merge without significant mass loss, one would expect this merger product to subsequently explode as a CCSN. Supernovae from stars modified by binary interacting could make up some 15 per cent fraction of all core-collapse supernovae, with a delay of $50$ to $200\,\text{Myr}$ relative to GC formation \citep{Zapartas2017}. Additionally, stripped massive stars, which are the progenitors of such supernovae, and their intermediate-mass counterparts, may provide a significant source of ionising radiation feedback in the cluster \citep{Stanway16,Gotberg19}, potentially contributing or even dominating the binary feedback channel. The main effect of delayed supernovae and stripped post-interaction stars would be to delay the start of the gas-rich epoch, where a gaseous reservoir can grow in the cluster core by the retention of stellar winds, and thus reducing, potentially significantly, the number of stars that can grow to become enriching stars. An important caveat here is that dynamical encounters within clusters may affect the binary evolution that produces the late CCSNe and post-interaction binaries. For example, dynamical disruption of soft binaries may be responsible for low binary fractions observed in globular clusters \citep{Milone12b} and partly responsible for the significantly lower binary fractions observed among the more concentrated second-population stars compared to first-population stars \citep{Lucatello15}. In addition to dynamical encounters, the binary stellar population in a gas-embedded region may be driven toward mergers \citep{Rozner2022} or the formation of contact binaries \citep{Menon2021} in which evolution towards equal masses in tight binary orbits can reduce the mass of the primary star and hence its ability to explode or produce a stripped star.

Thermonuclear supernovae (SNe Ia) are another potential significant source of energy \citep{Conroy2011, Milone2016}. Thought to be produced via mass transfer onto massive white dwarfs in binaries, or mergers of two white dwarfs, SNe Ia have typical field delay time distributions of $0.1$ to $1\,{\text{Gyr}}$, energy yields of $10^{51}\,\text{ergs}$, and production rates of the order $10^{-3}\,\text{M}_\odot^{-1}$ within a stellar population \citep{Maoz2014}. As with late CCSNe, the production rate of SNe Ia may be significantly reduced in GCs due to dynamical interactions or due to the presence of a large amount of gas within clusters. In this case, dynamical processes, dominated by stars more massive than $2
\,\text{M}_\odot$ will disrupt the binaries that form or contain WDs with typical masses below $1.4\,\text{M}_\odot$, preferentially ejecting WDs from the binaries. Indeed, \citet{Ivanova2006}, using Monte 
Carlo models of globular clusters, found that binaries containing WDs are produced inefficiently during the first Gyr. Finally, we acknowledge that detailed models of binaries in dynamical and gaseous environments are needed to confidently confirm that this feedback channel is indeed delayed within our model.

Accretion in interacting binaries containing WDs, NSs, and BHs could also provide significant feedback within GCs. Low-mass X-ray binaries (LMXBs) have a higher specific occurrence rate (i.e. number per unit stellar mass) in GCs than in the field \citep{Pooley03, Bregman2006} with many systems formed via dynamical encounters within clusters \citep[e.g.][]{DaviesBenz1995}. Persistent LMXBs can reach luminosities of about $10^{38}\,\text{erg}/\text{s}$, sufficient to unbind a $10^4\,\text{M}_\odot$ gas cloud within a cluster over a timescale of $10^5$ yr or so. LMXBs are likely to produce millisecond pulsars (MSPs). Winds from MSPs have also been proposed as a feedback mechanism for present-day globular clusters, capable of removing between $10^2$ and $10^3\,\text{M}_\odot$ of gas per $100\,\text{Myr}$ \citep{Spergel1991}. LMXBs and their MSP offspring are likely to be produced within globular clusters on timescales of Gyr \citep{Ivanova2008}, although the impact of gas on the production of these sources remains unknown.

Cataclysmic variables (CVs) have also been observed in large numbers within globular clusters \citep[e.g.][]{Haggard2009}. Novae, nuclear explosions on the surface of the accreting white dwarf in these systems, could also be an important source of feedback. Indeed, classical novae (CNe) from CVs have been proposed as the main source of feedback in present-day GCs \citep{Moore2011}. CNe have also been proposed as a possible source of chemical enrichment within clusters \citep{Maccarone2012}. It is unclear whether CV production is enhanced or reduced by dynamical interactions in clusters \citep{Davies97}, and there are currently no models of CV formation in gas-containing globular clusters. However, taking the field value for CNe rate would imply a rate of around $10^{-4}\,\text{yr}^{-1}$ in the most massive clusters, which would be sufficient to keep expelling all the gas in globular clusters today as each nova has an energy of about $10^{45}\,\text{erg}$ \citep{Moore2011}. Similar to LMXBs, a population of CVs will build up on the timescales of Gyr, but with many sources still active today \citep{Davies97}. 

In summary, we have demonstrated that disc crossings cannot continually remove the gas. Accretion onto compact objects, and especially massive compact objects, such as IMBHs, may serve as a significant source of feedback. However, it may be delayed or weakened due to birth kicks, accretion physics, beaming, or cooling processes in the GC gas. Core-collapse supernovae delay the gas epoch. Binary supernova progenitors are likely dynamically destroyed, although their impact remains a significant caveat for our model that requires further detailed simulations. Finally, accreting sources such as LMXBs and CVs are expected to be dynamically assembled on timescales of Gyr. Altogether, we acknowledge that feedback processes are uncertain and likely complex, and we further consider a plausible scenario in which feedback activates on timescales of approximately $1\,{\text{Gyr}}$.

\section{Time evolution of the gaseous reservoir}

\begin{figure}
    \centering
    \includegraphics{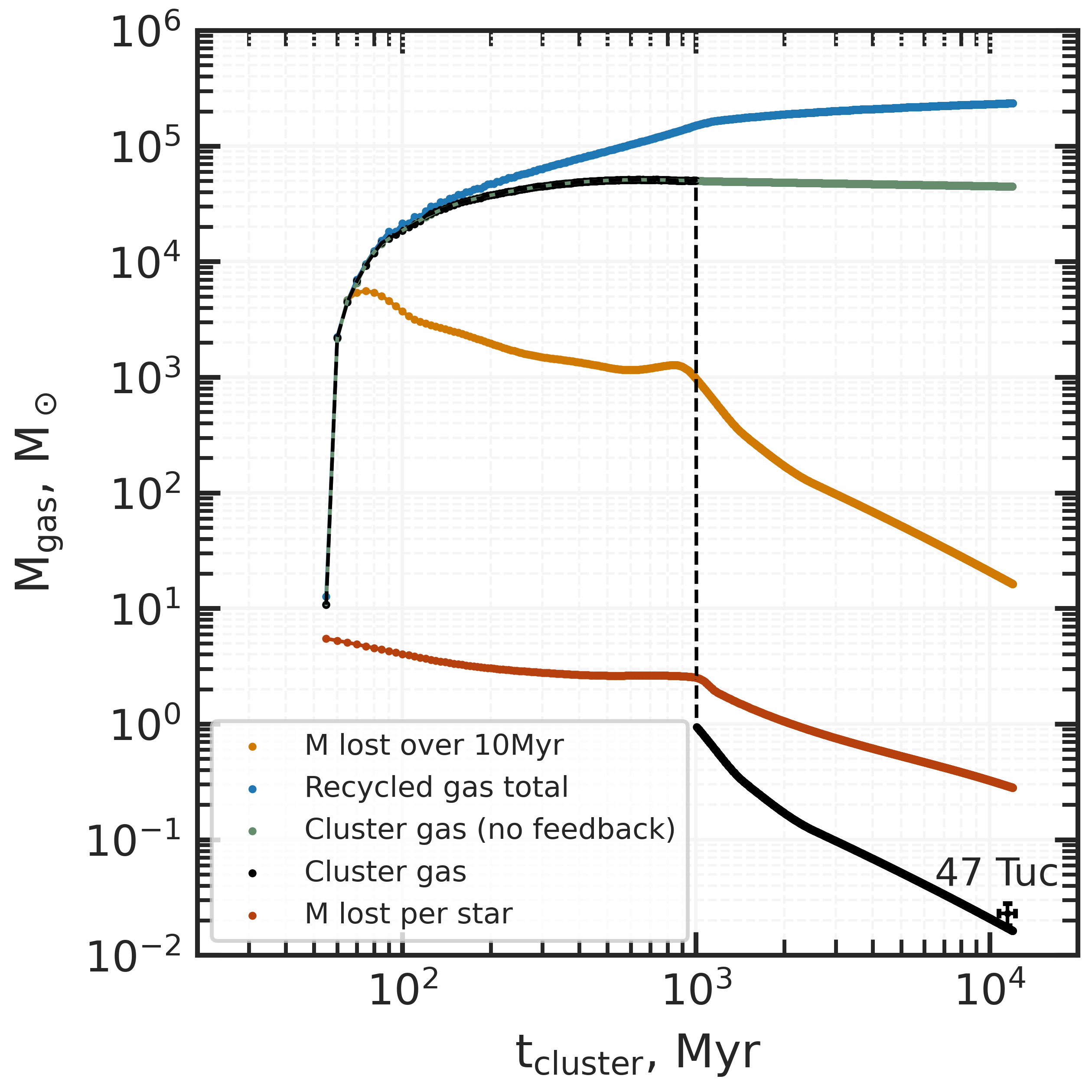}
    \caption{The evolution of cluster gas mass as a function of time, for a Kroupa population of $N=10^6$ stars at $[{\text{Fe}}/{\text{H}}]=-1.5$, as a function of cluster age. The blue line shows the cumulative total mass ejected by stars into winds. The red line shows the amount of mass injected by those individual stars undergoing high rates of mass loss at the end of their lives. This figure drops over time as the mass of stars losing mass decreases at later times, and their stellar evolution timescales increase. The orange line shows the total amount of gas injected by all stars during the preceding $10\,\text{Myr}$. One may see that a large fraction of the stellar mass loss via winds occurs in the first ${\text{Gyr}}$. The green line represents the mass of the gaseous reservoir at a given time, assuming gas depletion through Bondi-Hoyle accretion onto the stellar population, but excluding any feedback effects on the gas. Due to Bondi-Hoyle accretion, the amount of gas mass present in the cluster is at most about ten per cent of its total mass at all times. The black line represents the gas mass, assuming a feedback mechanism turns on at 1 Gyr. We assume that once feedback is turned on, only the gas emitted as winds in the previous 10 kyr is found in the cluster at any given time (following the suggested feedback mechanism of \citealt{Moore2011}, see Section~\ref{sec:Feedback} for details). For illustration, include a data point for the globular cluster 47 Tuc, which has an age of $12\pm 0.5\,\text{Gyr}$ \citep{Thompson2020} and which currently is essentially devoid of gas, with $M_{\text{gas}}=0.023\pm0.005\,\text{M}_\odot$ \citep{Abbate2018}. One can see that the feedback mechanism suggested by \citet{Moore2011} in combination with the present-day stellar wind production rate in the globular cluster is consistent with the observations.}
    \label{fig:MGasEvolution}
\end{figure}

\label{sec:GasEvol}

In Sections~\ref{sec:Winds} and \ref{sec:GasRet}, we showed that winds are effectively retained in globular clusters having sufficiently deep gravitational potential wells. Subsequently, in Section~\ref{sec:GasFate}, we argued that the evolution of the gaseous reservoir formed from the stellar-wind material is governed by its accretion onto stars and, later, by feedback processes. In this section, we introduce a time evolution model for this gaseous reservoir.

First, we consider the wind injection rate into the cluster from a Kroupa population of stars \citep{Kroupa2008} with a metallicity of $[\text{Fe/H}]=-1.5$. We initially bin and then sort the population by initial stellar masses $M_{\star,i}$ and obtain the wind injection rate for each mass from \citet{Choi2016}. The total wind injection rate from all stars is then given by:
\begin{equation}
    \dot{M}_{\text{wind, tot}} (t) = \sum_i \dot{M}_{\star,\text{wind},i} (t)
    \label{eq:StarWind}    
\end{equation}

Following the discussion in Section~\ref{sec:GasFate}, we assume depletion of the gaseous reservoir occurs through Bondi-Hoyle (B-H) accretion onto stars \citep{Bondi1944,Edgar2004}. Therefore, the rate of change of mass contained in the gaseous reservoir is given by:
\begin{equation}
    \dot{M}_{\text{gas}} (t) = \dot{M}_{\text{wind, tot}} (t) - \dot{M}_{\text{B-H, tot}}(t)
    \label{eq:mgasdot}    
\end{equation}
The total Bondi-Hoyle accretion rate is given by the sum of the individual accretion rates for each star:
\begin{equation}
    \dot{M}_{\text{B-H, tot}} (t) = \sum_i \dot{M}_{\text{B-H},i} (t)
    \label{eq:mbhdot}    
\end{equation}

As a result, the amount of gas follows an equilibrium defined by a balance between the retention of gas produced by stellar winds and the accretion of this gas onto pre-existing stars. This equilibrium value evolves on the timescale of stellar evolution. 

We further assume that the gas encloses a constant mass fraction $\alpha$ of all the cluster stars and maintains a uniform density $\rho_{\text{gas}}$ until removed by feedback processes. The gas mass is then related to density and radius through:
\begin{equation}
    M_{\text{gas}} (t) = \frac{4}{3}\pi \rho_{\text{gas}} (t) R_\alpha^3,
\end{equation}
with $R_\alpha$ defined in Equation~\ref{eq:RAlpha}. The B-H accretion rate for each star is then given by:
\begin{equation}
    \dot{M}_{\text{B-H},i} (t) = C (t) \rho_{\text{gas}} (t) M_{\star,i}^2,
    \label{eq:StarBH}
\end{equation} 
where $C_i \equiv {4\pi G^2}{(c_{\text{s}}^2 + v_i^2)^{-3/2}}$ is a constant for each star, independent of $\rho_{\text{gas}}$. Furthermore, for our purposes of modelling gas evolution, we assume that stars undergo B-H accretion only on the main sequence or as compact objects. Furthermore, we assume that stars are either always within the gaseous reservoir or never are.

When considering accreting stars on the main sequence, one needs to treat the ages of stars as more massive stars have much shorter main-sequence lifetimes. Therefore, it is possible to have at some instant rejuvenated stars with ages greater than the apparent lifetimes that could be inferred, e.g. from their photometry or current stellar masses. Here we allocate to each star their age in terms of their main-sequence lifetime fraction, $f_{\text{ms}}$, where stars initially have a value $f_{\text{ms}}=0$ and will evolve off the main sequence once $f_{\text{ms}}=1$. We then increment $f_{\text{ms}}$ each time step, using the current main-sequence lifetime, i.e., $\Delta f_{\text{ms}} = \Delta t / t_{\text{ms}}(m_i)$. We assume that the star will follow the ordinary post-main-sequence evolution for a star of its mass.

For a given cluster, one can numerically compute the mass of gas within the cluster as a function of time. For illustration, we consider a cluster containing one million stars initially, which evolve following MIST stellar evolution tracks with a metallicity [Fe/H]$= -1.5$. We assume that all gas retained in the cluster forms a sphere of uniform density with a radius containing initially one quarter of all stars. The cluster stars are taken to follow a Plummer model with a half-mass radius, $R_{\text{hm}}=2\,\text{pc}$ (which implies $R_{\text{gas}}=1.24\,\text{pc}$). We assume $R_{\text{hm}}$ and $R_{\text{gas}}$ do not evolve in time. We set the sound speed $c_{\text{s}} = 5\,\text{km}/\text{s}$. We do not model the density distribution of stars within the gaseous sphere, but consider them to all move at a constant speed $v_i = 10\,\text{km}/\text{s}$. Here, we do not consider accretion onto compact objects (which will be discussed in the subsequent section).

In Figure~\ref{fig:MGasEvolution}, we show the evolution of the gas mass. The total accumulated mass lost in stellar winds is shown as a blue line. For the cluster considered here, we would expect a large fraction of the mass to be retained (see Figure~\ref{fig:RetentionFractions}). One sees from Figure~\ref{fig:MGasEvolution} that most mass loss from stellar winds occurs in the first $1000\,\text{Myr}$ or so. The red line shows the mass lost by an individual star undergoing significant mass loss before becoming a compact object. Stars undergoing mass loss become less massive over time. Thus, the mass lost per star also goes down at later times. The orange line shows the total mass of gas injected in the preceding $10\,\text{Myr}$. One sees from Figure~\ref{fig:MGasEvolution} that the gas injection rate decreases over time, as stellar evolution timescales increase. The green line shows the mass contained within the gaseous reservoir derived assuming the wind-loss rates from the MIST stellar tracks and the analytic Bondi-Hoyle accretion rates onto the stellar population described above. No feedback is included. It is important to note that due to Bondi-Hoyle accretion, the amount of gas mass present in the gaseous sphere in the cluster centre at any given moment is at most about ten per cent of the total cluster mass at all times. This mass is much less than the accumulated mass of gas emitted as winds at later times.

The black line in Figure~\ref{fig:MGasEvolution} is the gas mass assuming a feedback mechanism turns on at a time of $1\,\text{Gyr}$. There are several potential sources of feedback, as discussed in Subsection~\ref{sec:Feedback}. We assume that once feedback is turned on, only the gas emitted as winds in the previous $10\,\text{kyr}$ is found in the cluster at any given time, following the suggested feedback mechanism of \citep{Moore2011}; see Subsection~\ref{sec:Feedback} for details. For illustration, we include a data point for the globular cluster 47 Tuc, which has an age of $12\pm 0.5\,\text{Gyr}$ \citep{Thompson2020} and which currently is essentially devoid of gas, with $M_{\text{gas}}=0.023\pm0.005\,\text{M}_\odot$ \citep{Abbate2018}. One can see that the combination of the feedback mechanism suggested by \citet{Moore2011} and the present-day stellar wind production rate in the GC are consistent with the observations.

\section{Possible consequences of gas retention}
\label{sec:Consequences} 

We have discussed earlier how gas from stellar winds retained in globular clusters will flow onto stars via Bondi-Hoyle accretion. In this section, we consider the effects of such accretion for three specific stellar populations: binaries, compact objects (i.e.\, black holes, white dwarfs and neutron stars), and main-sequence stars.

\subsection{Merger of stellar binaries}

Stellar binaries play a crucial role in the dynamical evolution of globular clusters.  As two-body scattering drains energy from the cluster core, encounters between binaries and passing single stars inject kinetic energy into the cluster, which helps offset core collapse. Some binaries are sufficiently wide that they are vulnerable to break-up when encountering passing stars and are termed soft. More tightly bound (hard) binaries become progressively harder through encounters and may ultimately be ejected from clusters or merge with other binaries. Thus, any primordial binary population is like a fossil fuel: it will be gradually used up over time.

Dynamical interactions between binaries and single stars are an important channel to produce various flavours of stellar exotica. A single neutron star may exchange into a binary which originally contained two main-sequence stars to produce a low-mass X-ray binary (LMXB). Roughly 10 per cent of persistent LMXBs are found in the Milky Way's globular clusters, whilst such clusters contain less than 0.1 per cent of the total stellar mass in the Galaxy \citep{Avakyan23}. A subsequent encounter between an LMXB and a second neutron star may produce a binary containing two neutron stars. If sufficiently tight, this binary will spiral in and merge via the action of gravitational radiation \citep{Ye20}. The same will apply for systems containing stellar-mass black holes \cite{Askar17}. Thus, globular clusters may be important sites for short gamma-ray bursts (thought to form from neutron star mergers) and gravitational-wave sources observed by ground-based detectors (merging systems contain stellar-mass black holes and/or neutron stars). 

The population of binaries within a globular cluster will be affected if they are engulfed in sufficiently dense gas \citep[see for example,][]{Rozner2022}. Gas may flow inward towards a binary, forming a circumbinary disc, which can then potentially extract angular momentum from the binary, causing it to shrink. Alternatively, gas drag on each of the two stars may cause them to lose momentum and thus cause the binary to shrink. Such shrinkage could remove an important energy source from the cluster (in terms of heating from binary-single encounters) and also enhance the merger rates of binaries, including those containing black holes and/or neutron stars \citep{Rozner2022}.

Binaries can be produced in globular clusters through tidal capture when two single stars pass sufficiently close \citep{Fabian1975}. The tidal capture rate may be enhanced by the presence of a gas which provides additional energy dissipation through gas drag \citep{Generozov2022,Rozner2023}. Further, the hard/soft boundary in the globular cluster population may be shifted as binaries, which would have been soft, i.e. ones which would have been broken up by encounters with single stars, and are first shrunk via the effects of gas drag, turning them into hard binaries,  before stellar encounters occur \citep{Rozner2024}.

\begin{figure}
    \centering
    \adjincludegraphics[width=\linewidth,trim={0 0 {.2\width} 0},clip]{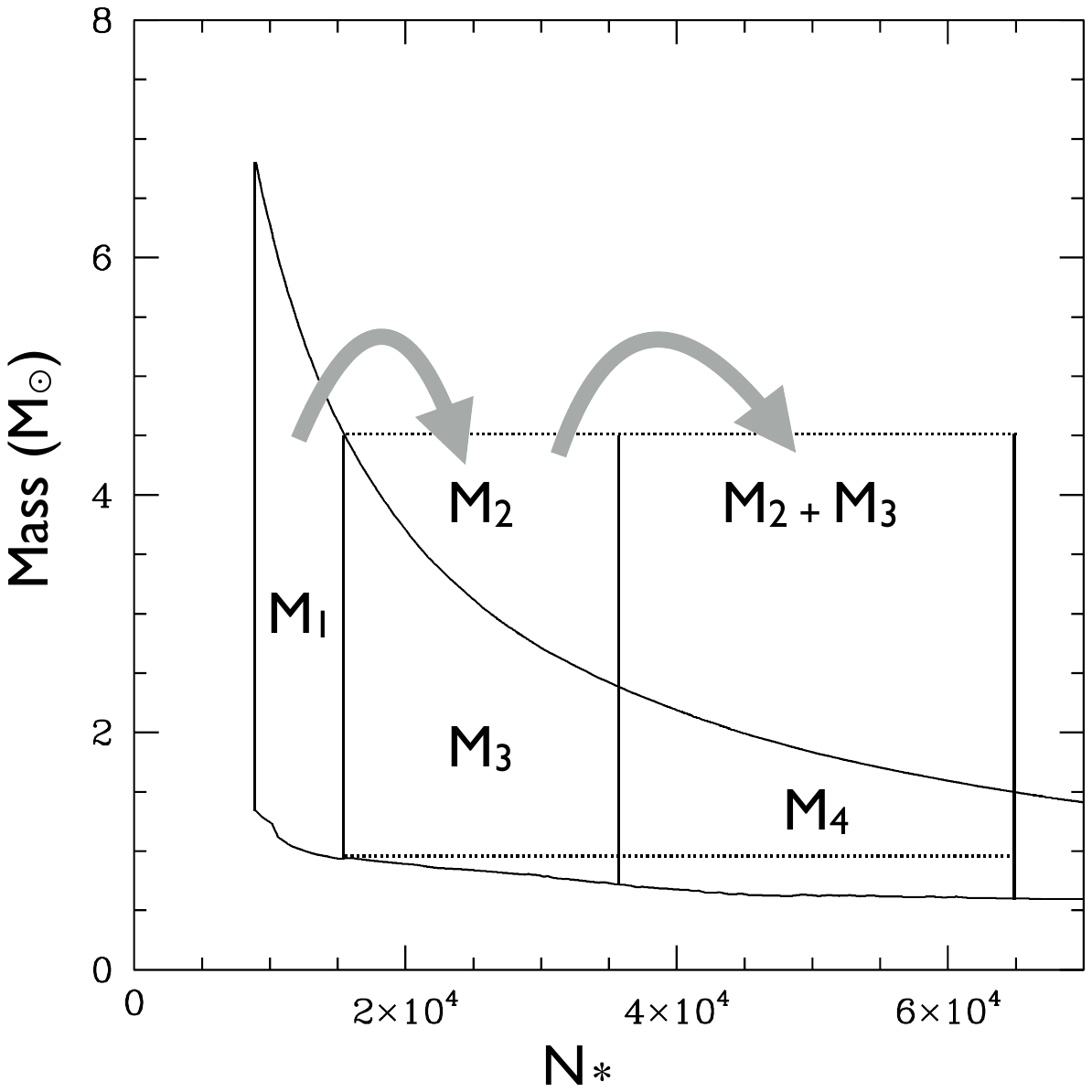}
    \caption{Diagram illustrating how mass effectively flows from the most massive stars to less massive stars, increasing their mass and in turn the number of stars above $4.5\,\text{M}_\odot$ which we term {\it enriching stars}. Here we have taken a population of $10^6$ stars following a Kroupa-Tout-Gilmore (KTG) IMF and ordered them in decreasing mass (i.e. most massive on the left). This population contains around $9000$ stars sufficiently massive to explode as SNe Type II ($M_\star > 6.8\,\text{M}_\odot$). We plot the initial masses of stars below this supernova limit vs their ranking $\text{N}_\star$. For a cluster of one million stars, $30000\,\text{M}_\odot$ or so of gas is produced from the original $6500$ enriching stars (labelled $M_1$ here). If this gas is optimally redistributed onto the next $21000$ stars, all of their masses would increase to $4.5\,\text{M}_\odot$. The total mass of (enriched) gas released by this second population of enriching stars amounts to about $75000\,\text{M}_\odot$ ($M_2 + M_3$, noting that $M_1=M_2$). If we were to repeat this procedure for a third population, the total mass of enriched material released in stellar winds is over $100000\,\text{M}_\odot$ ($M_2+M_3+M_4$).}
     \label{fig:mass_flow}
\end{figure} 

\begin{figure*}
    \centering
    \includegraphics[width=\textwidth]{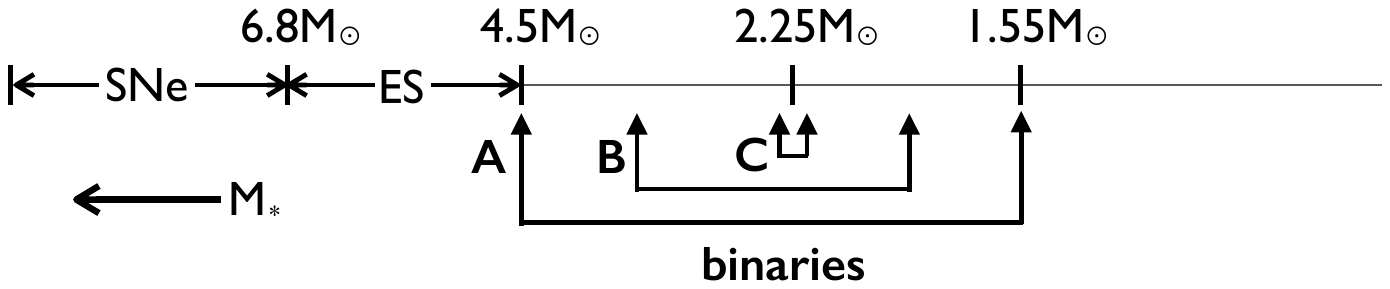}
    \vspace{-9.5truecm}
       \caption{A diagram illustrating how binary mergers may enhance
       the population of enriching stars (ES: $4.5\,\text{M}_\odot < M_\star < 6.8\,\text{M}_\odot$). Imagine them sorted in mass from left to right as shown above. For a KTG IMF, there are an equal number of stars between $2.25\,\text{M}_\odot$ and $4.5\,\text{M}_\odot$ and between $1.55\,\text{M}_\odot$ and $2.25\,\text{M}_\odot$. Therefore, one may optimally arrange stars in binaries as shown above with binaries A, B, and C all containing a total stellar mass of $4.5\,\text{M}_\odot$. Thus, if the presence of a gaseous reservoir in a globular cluster were to force the binary components to merge, each merger would produce a $4.5\,\text{M}_\odot$ (enriching) star.}
     \label{fig:merging_binaries}
\end{figure*}

\subsection{Growth of compact objects}

Globular clusters are observed to contain far more LMXBs and millisecond pulsars per unit stellar mass compared to the field \citep[e.g.,][]{Pooley03}. It is thought these populations are produced via dynamical encounters between single neutron stars and binaries containing main-sequence stars \citep[e.g.,][]{Davies98}. A globular cluster is likely to produce between $3000$ and $10000$ neutron stars in core-collapse supernovae. However, neutron stars are observed to receive large natal kicks at birth \citep{Hobbs2005,Disberg25}. Such kicks are likely to eject the neutron stars from globular clusters. However, the observed X-ray population in globular clusters would seem to require that at least $10$~--~$20$ per cent of neutron stars are retained \citep{Pfahl2002}. Binary companions may help in retaining neutron stars, but are unlikely to be the complete answer \citep{Davies98}.

\citet{Ivanova2008} have suggested that neutron stars produced in electron capture supernovae may receive much smaller natal kicks and thus be retained in globular clusters. This channel could produce a population equivalent to a retention fraction of $10$ per cent. A more recent, and related, suggestion \citep{Perets2022} is that massive ONe white dwarfs could be induced to collapse to form neutron stars through the accretion of additional material from the surrounding gas; a process known as accretion-induced collapse or AIC \citep[e.g.\,][]{Nomoto1991}. AIC may occur within binaries, but also when a white dwarf is embedded in a gaseous reservoir, as discussed in \citet{Perets2022}. It is also important to note that accretion from a gaseous reservoir onto CO white dwarfs could produce type Ia supernovae, e.g., \citet{Ruiter25}. The energy released in such supernovae could act as an important feedback mechanism, potentially ejecting the gaseous reservoir from the stellar cluster. We will return to this in a subsequent paper in the series.

The accretion rate is a critical parameter in determining whether accreting white dwarfs grow in mass. This is because nuclear explosions on its surface (novae) may, in fact, eject a large fraction of the accreted material. In some cases, the white dwarf mass can even decline \citep[e.g.\,][]{Yaron2005}. In order to avoid too much mass loss, a high accretion rate needs to take place, which in turn implies relatively cold and dense gas (as can be seen from Equation~\ref{eq:BH1}). If we assume all ONe white dwarfs with masses above $1.2\,\text{M}_\odot$ can form neutron stars through AIC, then a cluster containing one million stars may well produce around $1000$ neutron stars in this way. If a newly formed neutron star remains in a gas cloud, further accretion may cause it to grow beyond the maximum possible mass for a neutron star, and thus it will turn into a stellar-mass black hole. 

We return to the stellar cluster evolution used to produce Figure~\ref{fig:MGasEvolution}, but this time we include WD accretion, assuming an accretion efficiency of $0.3$, i.e. some $30$ per cent of the material flowing onto a WD via Bondi-Hoyle accretion is retained. Further, we consider that WDs collapse to form neutron stars once they exceed $1.4\,\text{M}_\odot$ and that neutron stars collapse to form black holes once they exceed $2.4\,\text{M}_\odot$. We assume both neutron stars and black holes accrete $100$ per cent of the material acquired through Bondi-Hoyle accretion. NS and BH accretion is not allowed to exceed the Eddington accretion rate. In practice, the accretion rates are not seen to exceed 10 per cent of the Eddington rate. We find that in the first $1000\,\text{Myr}$, massive white dwarfs derived from stars above $5.84\,\text{M}_\odot$ are found to grow sufficiently to become at least neutron stars. Indeed, a large fraction of this population continues to grow, becoming black holes, on the assumption that they remain in the core (and embedded in the gaseous reservoir) at all times.

Finally, in this section, we consider the growth of black holes by accretion of gas, including both the black holes produced through the growth of white dwarfs (becoming neutron stars first, then black holes), but also any black holes produced through the core-collapse (Type II) supernovae of the most massive stars. Indeed, accretion onto the second group of black holes could lead to the depletion of gas left over from the first phase of star formation, perhaps even before a substantial wave of SNe Type II from stars producing neutron stars ejects the remaining gas \citep{Leigh2013}. 

For both groups of black holes, it is possible, given favourable gas densities, that black holes reach masses significantly larger than those produced in SNe Type II. Such black holes may eventually dominate B-H accretion in the cluster and quench the gas-rich epoch. As has been noted by \citet{Roupas2019b,Roupas2019a} and others, the production of these more massive black holes might well explain some of the systems containing massive black holes observed as gravitational wave sources by LIGO.

\subsection{Rejuvenation of main-sequence stars}

As discussed earlier, most globular clusters are found to possess at least two stellar populations \citep[][and references contained therein]{Bastian2018}. The two stellar populations are often found to have differences in oxygen and sodium abundances, with the second population possessing a larger sodium abundance and smaller oxygen abundance than the first population. Stellar winds from intermediate-mass stars (stars between about $4.5\,\text{M}_\odot$ and $7\,\text{M}_\odot$) may be sodium-rich and oxygen-poor due to nuclear burning reactions taking place within them \citep[e.g.][]{DAntona2016}. We term such stars {\it enriching stars} from now on. One possible interpretation of the observations of multiple populations in globular clusters is that winds from enriching stars have polluted a subset of the first population to produce the second population. It is important to realise that this is distinct from the idea that the second population is produced via star formation in gas accumulated from stellar winds \citep[e.g.][]{DErcole2008}; earlier in Subsection~\ref{sec:StarForm}, we saw how star formation within a gaseous reservoir is prevented because of encounters between contracting cores of gas (which would otherwise proceed to form stars) and pre-existing stellar cluster stars.

The observed abundances of the second population are midway between those of the first population and those expected for enriching star winds \citep[e.g.\,][]{Bastian2018}. As stars are likely to be well mixed, this implies that a significant fraction of the gas in a second population star (between $20$ and $50$ per cent) has been accreted from the winds of enriching stars.

If we consider a cluster of one million stars, drawn from a Kroupa IMF, then some $980000$ stars will be less massive than $4.5\,\text{M}_\odot$, with a total mass of about $500000\,\text{M}_\odot$. For comparison, the total stellar wind mass for the stars having initial masses between 4.5 M$_\odot$ and 7 M$_\odot$ for the same stellar cluster would be only about $30000\,\text{M}_\odot$. In other words, only a small fraction of the mass of the stars we wish to pollute. However, the total stellar mass is reduced somewhat when we consider only the $650000$ stars which are less massive than $0.4\,\text{M}_\odot$. These have a total mass of about $130000\,\text{M}_\odot$. If we were to double the mass of $0.4\,\text{M}_\odot$ stars in the first population, they would produce the most massive second population stars observed today (as globular cluster turn-off masses are around $0.8\,\text{M}_\odot$). Furthermore, three effects could considerably enhance the production of enriched gas, as described below.

Firstly, it is important to recall that globular clusters have undergone considerable mass loss since their formation. Tidal losses are a function of a cluster's location and orbit within the Milky Way, but in many cases, clusters have lost around two-thirds of their stars \citep{Baumgardt2018}. The stars which are tidally stripped away from a cluster will come from the outer regions of the cluster, with mass loss timescales being comparable to the cluster age. Therefore, if retained stellar winds settle into the central regions, containing perhaps around a quarter of the original stars, then these stars will be enriched to form a second population, whilst a considerable fraction of the outer (first) population is removed. Cluster mass loss on the timescale of the original intermediate-mass stars may be very low. Hence, this way we may boost the production of enriched gas (per second population star) by factors of $3$ to $5$. 

Secondly, the population of enriching stars will itself be boosted by Bondi-Hoyle accretion onto lower-mass stars. For example, accretion of $1\,\text{M}_\odot$ of gas onto a $4\,\text{M}_\odot$ star will produce a new enriching star. Recall also how in Bondi-Hoyle accretion, more massive stars accrete a much larger amount of gas than lower-mass stars in the same environment. One can therefore imagine a rolling wave of accretion where, at any moment, the most massive stars are accreting most of the gas until they evolve into compact object remnants, releasing the (often enriched) stellar winds for subsequent re-accretion. To illustrate how this may increase the total mass of enriched material available, let us consider the following idealised picture. Imagine all stars within a cluster are single and sort them in order of decreasing mass, as illustrated in Figure~\ref{fig:mass_flow}. Consider the stellar winds released by enriching stars to accrete in an optimised way onto stars just below $4.5\,\text{M}_\odot$ such that all stars increase in mass, such that they just become enriching stars themselves (i.e., they accrete until they have masses of $4.5\,\text{M}_\odot$). For a cluster of one million stars, the $30000\,\text{M}_\odot$ or so of gas from the original $6500$ enriching stars would accrete onto some $21000$ stars, increasing all their masses to $4.5\,\text{M}_\odot$. The total mass of (enriched) gas released by this second population of enriching stars amounts to about $75000\,\text{M}_\odot$. If we were to repeat this procedure for a third population, the total mass of enriched material released in stellar winds is over $100000\,\text{M}_\odot$.
 
We can achieve a similar result by placing stars in binaries. As discussed earlier, the presence of a gaseous reservoir will cause many binaries to merge. One could then, for example, produce a single enriching star by merging a $2\,\text{M}_\odot$ star and a $3\,\text{M}_\odot$ star in a binary. How much enriched material could one produce through the merger of lower-mass stars in binaries? To answer this, we consider the optimised arrangement of stars in binaries as shown in Figure~\ref{fig:merging_binaries}. Sorting stars in order of decreasing mass, a cluster containing one million stars, will possess some $20000$ or so stars having masses between $2.25\,\text{M}_\odot$ and $4.5\,\text{M}_\odot$. Pairing these up with lower mass stars, as illustrated in Figure~\ref{fig:merging_binaries}, produces some $20000$ binaries where the total stellar mass in each exceeds the $4.5\,\text{M}_\odot$ required to produce an enriching star. Let us assume that all binaries merge to produce single stars. In such a case, the total mass of enriched material released would be around $100000\,\text{M}_\odot$.

The gas released by stars after the start of the feedback-dominated epoch will likely not accrete on the cluster stars significantly, since the equilibrium amount of gas in the cluster will remain low. Therefore, the enriched material contained in stars more massive than the present-day turnoff stars, approximately $0.8\,\text{M}_\odot$ will eventually be lost from the cluster. Additionally, some stars will have accreted unenriched material after accreting enriched material. In this case, the enriched material may appear on the surface of the star after the mixing timescale within the star, potentially Gyrs later, which may have to be accounted for when interpreting young globular cluster populations.

\begin{figure}
    \centering
    \adjincludegraphics[width=\linewidth,trim={0 0 {.2\width} 0},clip]{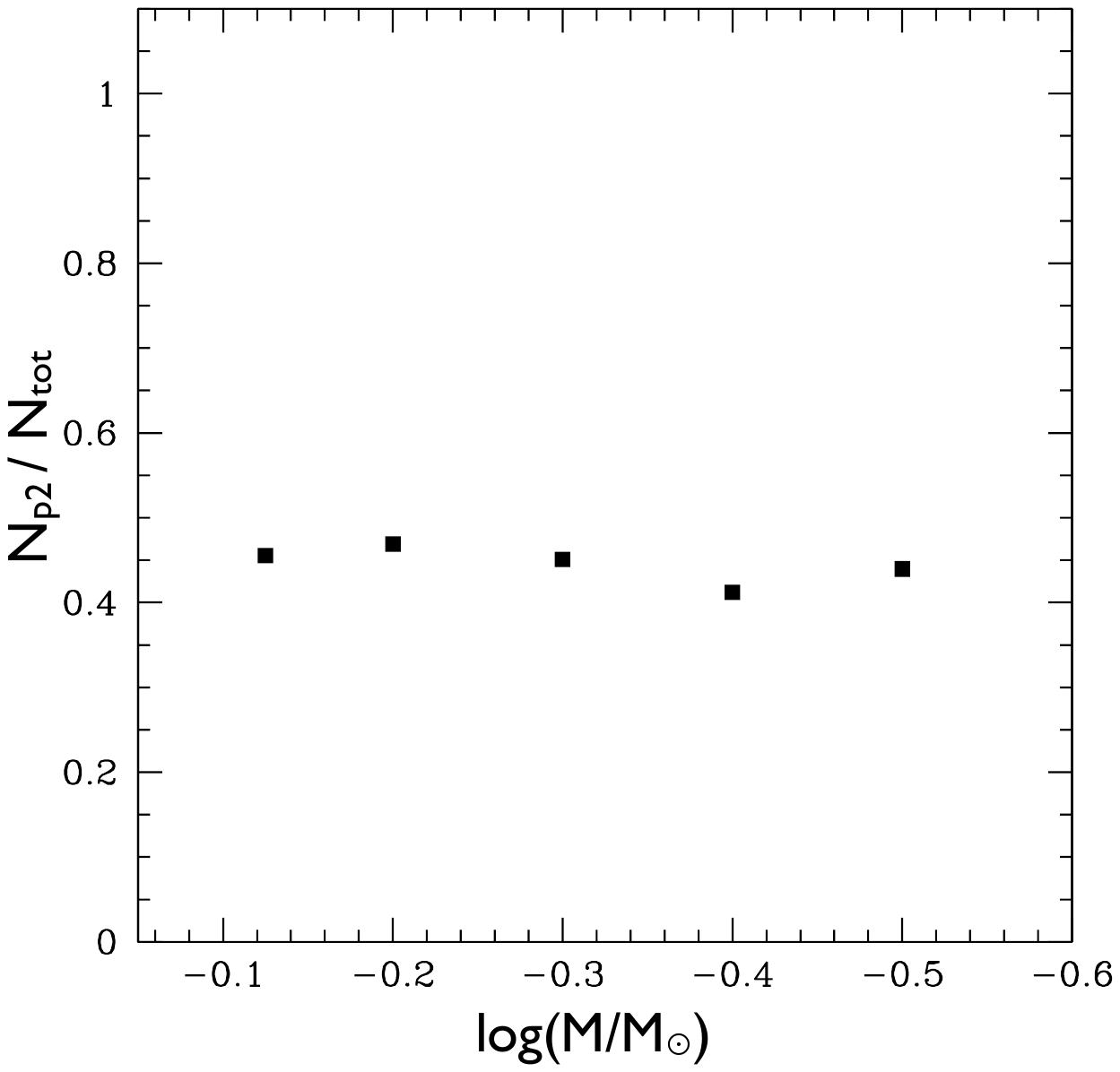}
       \caption{A plot showing the fraction of stars belonging to the second population (P2) as a function of log(mass), where the second population is produced via Bondi-Hoyle accretion onto the first population. The key point here is that the fraction is relatively flat (as is observed), meaning that Bondi-Hoyle accretion may preserve the stellar mass function.}
     \label{fig:ratio_plot}
\end{figure} 

Finally, before we close this section, we consider the distribution of stellar masses for both the first and second populations in a globular cluster. Observations of NGC 2808 and NGC 6121 (M4) reveal that the fraction of stars drawn from the second population is independent of mass over a broad range of masses \citep{Dondoglio22}. This observation would seem to challenge the idea that the second population is derived from the first via a phase of significant mass growth due to Bondi-Hoyle accretion. As the Bondi-Hoyle accretion rate varies significantly with stellar mass, one might have expected such accretion to lead to a different distribution of stellar masses. However, intuition can sometimes mislead. In fact, {\it Bondi-Hoyle accretion is likely to produce a second stellar population having a mass distribution very similar to that of the first population}, as we explain below.

As we saw earlier in Equation~\ref{eq:BH2}, a star of initial mass $M$ reaches a mass $\tilde{M}$ after some time $\Delta t$, where $K$ is the constant associated with the Bondi-Hoyle accretion rate, where $\tilde{M}$ is given by
\begin{equation}
\label{eq:BHMod}
\tilde{M} = \frac{M}{1 - K \Delta t M}
\end{equation}

One can use the above equation to calculate the final masses for a population of stars. We consider stars following the Kroupa IMF with masses between $0.1\,\text{M}_\odot$ and $0.4\,\text{M}_\odot$. We consider that a subset of the stars that accrete gas to become a second population and scale the accretion such that $0.4\,\text{M}_\odot$ stars accrete a further $0.4\,\text{M}_\odot$, i.e. $\tilde{M} = 0.8\,\text{M}_\odot$. Binning the enriched stars in log mass, we plot the fraction of second population as a function of log mass in Figure~\ref{fig:ratio_plot} and find that the fraction is flat, reproducing the observations reported in \citet{Dondoglio22}.

One can also reach a similar result through an analytic treatment. Rearrangement of Equation~\ref{eq:BHMod} provides an expression for $M$ in terms of $\tilde{M}$ 
\begin{equation}
M = \frac{\tilde{M}}{1 + K \Delta t \tilde{M}}
\end{equation}

The IMF, $dN/dM = f(M)$. One can derive the new IMF for the stars post-accretion using the chain rule
\begin{equation}
\frac{dN}{d\tilde{M}} = \frac{dN}{dM} \cdot \frac{dM}{d\tilde{M}} = f(M) \left[ \frac{1}{1 + K \Delta t \tilde{M}}
- \frac{K \Delta t \tilde{M}}{(1 + K \Delta t \tilde{M})^2} \right]
\end{equation} 
The IMF can therefore be written as a function of $\tilde{M}$
\begin{equation}
\frac{dN}{d\tilde{M}} =  \tilde{f}(\tilde{M}) = f\left(\frac{\tilde{M}}{1 + K \Delta t \tilde{M}}\right) \cdot
\frac{1}{(1 + K \Delta t \tilde{M})^2}
\end{equation} 
If $f(M) = C M^{-\gamma}$, then we have
\begin{equation}
\tilde{f}(\tilde{M}) = C \tilde{M}^{-\gamma} \cdot (1 + K \Delta t \tilde{M})^{\gamma -2 }
\end{equation} 
which implies $\tilde{f}(\tilde{M}) = f(M)$ for $\gamma = 2$. In other words, Bondi-Hoyle accretion following Equation~\ref{eq:BHMod} will act in such a way to preserve the stellar mass function for $\gamma = 2$, which is indeed close to the power law observed in stellar clusters. Finally, we note that these considerations apply to populations with velocities not too different due to equipartition, which is a good approximation for dynamically young globular clusters \citep{Bianchini16, Watkins22}.

\section{Summary}

\label{sec:Summary} 

In this paper, we have considered the retention of gas within globular clusters. In particular, considering the fate of stellar winds. Our key findings are listed below:

\begin{enumerate}
    \item Gas emitted as stellar winds is likely to be retained in globular clusters but not in open clusters, as the wind speeds are somewhat larger than open cluster escape speeds, but somewhat smaller than globular cluster escape speeds (i.e.\,$v_{\text{esc,oc}} \ll v_{\text{w}} \lesssim v_{\text{esc,gc}}$). Furthermore, stars experience most of their mass loss through stellar winds in a relatively short phase. Only approximately one per $10^5$ stars will be in this phase at a given time. In clusters containing more than $\sim 10^5$ stars, collisions between winds will further help in gas retention. The mass contained in the gaseous reservoir, $M_{\text{gas}}$, will greatly exceed the mass emitted as a wind by an individual star (i.e. $M_{\text{gas}} \gg M_{\star,\text{wind},i}$). Thus, once gas begins to be retained, the reservoir will form a blanket, greatly enhancing gas retention.
    \item The encounter time between a cooling core of gas within the gaseous reservoir and pre-existing stars within the cluster is much shorter than the star formation timescale within the core (i.e. $t_{\text{enc}}({\text{core}}-\star) \ll t_{\text{form}}({\text{core}})$) hence star formation within the gas will be inhibited. Rather, accretion onto pre-existing stars is more likely to occur.
    \item Stars will therefore grow via Bondi-Hoyle accretion, the rate is such that lower-mass stars are likely to grow up to masses around $4.5$~--~$6\,\text{M}_\odot$ thus increasing the number of stars which will produce chemically-enriched stellar winds which may in turn be useful in producing the chemical enrichment seen in the second stellar population in globular clusters. We therefore term stars with masses $4.5\,\text{M}_\odot < M_\star < 6.8\,\text{M}_\odot$ {\it enriching stars}. An equilibrium is reached between mass being added to the reservoir via stellar winds and its subsequent re-accretion onto other stars, such that the mass contained in the reservoir is much less than the total amount of gas emitted as stellar winds (i.e.\,$M_{\text{gas}} \ll \Sigma M_{\text{wind},\star}$). Because higher-mass stars will accrete at much higher rates than lower-mass stars, a large fraction of the gas will accrete onto the most-massive stars at a given time. These stars, in turn, evolve and re-emit the gas as stellar winds. Thus, we will have a {\it stellar conveyor belt} where the gas is effectively passed on from the most massive to the least massive stars.
    \item Other work \citep[e.g.\,][]{Rozner2022} has shown that the presence of a gaseous reservoir may considerably shorten the merger timescale for binaries. With this in mind, we have shown here that such binary mergers may also significantly increase the population of enriching stars. 
    \item Our calculations here suggest that one can potentially produce the second population of stars observed in globular clusters by Bondi-Hoyle accretion onto pre-existing first populations. Such a population may preferentially be formed in cluster centres (where the gaseous reservoir is likely to form). We will consider how accretion shapes the stellar population in the second paper in this series (Bobrick et al., in prep).
    \item It is important to note that the stellar mass functions observed in globular clusters are (perhaps surprisingly) invariant under Bondi-Hoyle accretion.
    \item Gas from the reservoir is also likely to accrete on any compact objects (white dwarfs, black holes and neutron stars) located within the reservoir. Our calculations here suggest that indeed white dwarfs can accrete sufficient mass to become neutron stars as suggested by \citet{Perets2022}. It is possible that some neutron stars, in turn, grow to become black holes. These black holes, together with any formed via core-collapse supernovae in massive stars, may accrete significant amounts of gas, potentially reaching intermediate mass scales ($\sim 100$~--~$300\,\text{M}_\odot$) and also explaining the massive black hole systems observed by LIGO.
    \item We expect feedback mechanisms to turn on at some point around $1\,\text{Gyr}$, which will eject the gaseous reservoir and prevent the subsequent retention of gas from stellar winds, for example, energy released as gas is accreted onto compact objects. Our picture is therefore consistent with observations of globular clusters today, which are found to contain an extremely small amount of gas. 
\end{enumerate}

\section*{Acknowledgements}

We thank Abbas Askar, Holger Baumgardt, Aleksey Generozov, Mor Rozner, Andrew Winter and Claire Ye for valuable discussions. We also thank Holger Baumgardt for kindly providing the most recent values for the initial masses of the globular clusters in the LMC and SMC through private communication and the anonymous referee for their valuable comments. A.B.~acknowledges support from the Australian Research Council (ARC) Centre of Excellence for Gravitational Wave Discovery (OzGrav), through project number CE230100016. A.B. and H.B.P.~acknowledge support for this project from the European Union's Horizon 2020 research and innovation program under grant agreement No 865932-ERC-SNeX.

\section*{Data availability}

The data underlying this article will be shared on reasonable request to the corresponding authors.

\bibliographystyle{mnras}
\bibliography{main}

\bsp	
\label{lastpage}
\end{document}